\documentclass[useAMS,usenatbib]{mn2e}

\usepackage[latin1]{inputenc}
\usepackage[T1]{fontenc}
\usepackage[dvips]{graphicx}

\usepackage{graphicx}
\usepackage{amssymb}
\usepackage{subfigure}

\title[STELLAR METALLIC ENHANCEMENT BY A PLANETESIMAL BOMBARDMENT]{A POSSIBLE STELLAR METALLIC ENHANCEMENT IN POST-T TAURI STARS BY A PLANETESIMAL BOMBARDMENT}
\author[O. C. Winter and R. de la Reza and R.C. Domingos and L.A.G. Boldrin and C. Chavero] {O. C.
Winter$^{1}$\thanks{ocwinter@feg.unesp.br}, R. de la Reza$^{2}$,
R. C. Domingos$^{1}$, L. A. G. Boldrin$^{1}$ and C. Chavero$^{2}$\\
$^{1}$S\~ ao Paulo State University - UNESP, Grupo de Din\^{a}mica Orbital \& Planetologia, Guaratinguet\'{a}, CP 205, CEP 12.516-410, Brazil.\\
$^{2}$Observat\' orio Nacional, Rio de Janeiro, RJ, Brazil}

\begin{document}
\maketitle

\label{firstpage}

\begin{abstract}
The photospheres of stars hosting planets have larger metallicity than stars lacking planets.
This could be the result of a metallic star contamination produced by the bombarding of hydrogen deficient solid bodies. In the present work we study the possibility of an earlier metal enrichment of the photospheres by means of impacting planetesimals during the first 20-30Myr. Here we explore this contamination process by simulating the interactions of an inward migrating planet with a disc of planetesimal interior to its orbit. The results show the percentage of planetesimals that fall on the star. We identified the dependence of the planet's eccentricity ($e_p$) and time scale of migration ($\tau$) on the rate of infalling planetesimals. 
For very fast migrations ($\tau=10^2$yr and $\tau=10^3$yr) there is no capture in mean motion resonances, independently of the value of $e_p$. Then, due to the planet's migration the planetesimals suffer close approaches with the planet and more than 80\% of them are ejected from the system.
For slow migrations ($\tau=10^5$yr and $\tau=10^6$yr) the percentage of collisions with the planet decrease with the increase of the planet's eccentricity. For $e_p=0$ and $e_p=0.1$ most of the planetesimals were captured in the 2:1 resonance and more than 65\% of them collided with the star.
Whereas migration of a Jupiter mass planet to very
short pericentric distances requires unrealistic high disc masses, these
requirements are much smaller for smaller migrating planets. Our simulations
for a slowly migrating 0.1 $M_{\rm Jupiter}$ planet, even demanding a possible
primitive disc three times more massive than a primitive solar nebula, produces maximum
[Fe/H] enrichments of the order of 0.18 dex. These calculations open
possibilities to explain hot Jupiters exoplanets metallicities.

\end{abstract}


\section{Introduction}

A particular interesting problem concerning exoplanetary science is
metallicity. Even if the near 200 extrasolar giant planets that have
been detected by transit and mainly radial velocity methods are
concentrated around
in general nearby solar type stars, a clear metallicity correlation
has been established. In fact, stars with planets (SWP) appear to be
more metal rich than stars without planets. A general metallicity
shift of ~0.2 dex in
[Fe/H] is characteristic for SWP (Santos et al. 2005). In particular
it is interesting to note that stars with hot Jupiters (with orbital
periods less than about 5 days) appear to be more metal rich than the
mean of all SWP (Sozzetti et al. 2004; Fischer \& Valenti 2005; Butler
et al. 2006). These results are concentrated in surveys and analyses
of non-metal poor stars and for spectral types F (young stars), G and K
(old stars). Two important review articles devoted to the chemical composition
of SWP, indicating also properties of other elements than Iron, appeared
with results obtained up to 2003 (Gonzalez 2003) and for the period
2003-2006 in Gonzalez (2006).

What is the physical mechanism behind these metallicity relations ?
Two main different interpretations have been proposed in the
literature, nevertheless, any of them has produced a satisfying
explanation. On the one hand, is a primordial condition in which metal
rich SWP would have been formed in metal rich clouds. On the other
hand, there is the external accretion mechanism. Here the stellar
surfaces, containing poor or normal metal abundances enhance their
metallicities by the injection of solid metal rich material depleted
in H and He.

These two mechanisms will form, metallically speaking, two different types
of stars. The initial or primordial mechanism, produce entire metal
rich stars, from their centres up to their atmospherical layers. This
is not the case for the accretion scenario, which can modify the metal
content of the external layers only, at least for those stars with
shallow convective
layers. The primordial mechanism is only considered in the literature
when any argument in favor of the self-enrichment is discarded. Anyway,
the primordial mechanism necessitates a highly inhomogeneous
metallicity distribution of the
interstellar matter, not detected at present. A powerful argument in
favor of the primordial case, would consist in finding that the
centres of SWP by asteroseismological methods, are also metal rich.
The only star  studied with this technique was $\mu$ Arae 
(Bazot \& Vauclair, 2004; Bazot et al. 2005).

Independent of this, the accretion or self-enrichment mechanism can
continue to be explored and this is the purpose of the present work.
We study here the possibilities of bombardment of stars atmospheres by
their discs planetesimals during a certain phase of the pre-main
sequence (PMS) evolution. We hope that this will provide some clues
respect to metallic enrichment.

In the next section we discuss some aspects of the evolution of PMS discs. The
numerical model is presented in Section 3. The results from our
simulations are given in section 4. In Section 5 some problems related
to the mass of discs are discussed. Section 6 is devoted to the metal
enhancements and finally in Section 7 we present our conclusions.

\section{Pre-Main Sequence Disc Evolution}

Several stages, some of them badly known, are involved in discs
evolution in the PMS. If discs during the T Tauri phase (before 5 Myr)
are formed by a relative homogeneous distribution of gas and dust,
during the next post- T Tauri phase the situation is very different.
In fact, in post- T Tauri stars (PTTS) disc gas and dust components
follow quite different histories. Gas is lost in a canonical period of
~10 Myr (see for instance Sicilia-Aguilar et al. 2005).
However. PTTS associations belonging to the older subgroups of the
Sco-Cen OB association appear to retain their disc gas up to ages of
~16 Myr (Pinz\'on et al. 2007). In any case, during the first 30 Myr,
the fine dust content is gradually agglutinated into kilometric sized
rocky or iced bodies (planetesimals). Eventually, in the outer parts
of the discs, conditions are fulfilled for the formation of a core of
10 - 20 $M_{\rm Earth}$ capable to form a giant planet, attracting the gas when
this is still available.

When gas is almost absent, dust evolution can be more complex because
energetic collisions between planetesimals, produce a new generation fine dust
forming what is called a Debris Disc (DD) which can be maintained up to very
large ages, even Gyr. At present, our knowledge of planetary
formation in DD is confused.

Considering the above discussion, it is possible to conceive the existence of
discs with ages between 10 and 30 Myr in which an external planet
coexist with a sea of internal planetesimals. If the total mass of
planetesimals is of the same order of the planet, interactions between
these two components will provoke an internal migration of the planet.
This because the planet is continously loosing energy and the excess
energy is being used to disperse the planetesimals (Murray et al.
1998). Since their work, planetary migration has become an important
part in the study of planetary systems. Some recent lectures or
reviews on this, are available in the literature. For extra solar
systems in Artimowicz (2006) and for the solar system in Levison et
al. (2007). More particularly, several recent studies appeared using
N-body simulations in order to study the conditions of survival (or
formation) of terrestrial type planets due to migration of a giant
planet (Lufkin, Richardson \& Mundy 2006; Fogg \& Nelson 2005; Raymond
et al. 2006; Armitage 2003; Mandell \& Sigurdsson 2003). In respect to
N-body simulations of
planetesimals and a migrating planet directed to the metallicity
enhancement problem, the only work to our knowledge is that of Quillen
\& Holman (2000) (hereafter QH00). In principle our work here is an
extension of QH00.

\section{Numerical Model}

In this work, we numerically integrated planar systems composed by a star, a planet
and one thousand planetesimals. It was assumed that the star has the same mass as
our Sun and the planet the same mass as Jupiter.
The planet is initially with semi-major axis $a_p=5$AU and eccentricity
$e_p$. The planet is forced to migrate inward up to $a_p=0.01$AU, with constant speed
in a timescale $\tau$.
The planetesimals are considered to be massless particles that are initially distributed
on random circular orbits with semi-major axis $1<a<4$AU.
It is important to note that this migration is not self-consistent in that
the bodies ejected do not cause migration.

The numerical simulations were performed using a code from SWIFT (Levison \& Duncan, 1994).
The integrator is based on the MVS method developed by the Wisdom \& Holman (1991).
The output of the simulations give the temporal evolution of the semi-major axis and
eccentricity of the planetesimals. The integration for each particle stopped when
one of the three following conditions happened:

a - {\sl collision with the star}: when the planetesimal gets closer than 0.01AU from the star;

b - {\sl collision with the planet}: when the planetesimal gets closer than two Jupiter's radius from the planet;

c - {\sl ejection from the system}: when the planetesimal reaches more than 50AU from the star.

The value of the time scale, $\tau$, is defined by the process responsible for the planet's migration Two main process are usually discussed in the literature. One is the migration of the planet due to the gravitational interaction with a disc of planetesimals (Murray et al. 1998). The other is the migration of a planet embedded in a gaseous disc (Artymowicz 2005). In the first case the time scale is of the order of $10^5-10^7$ yr, while in the second case it is much faster, $10^2-10^3$ yr. In our studies we tried to cover the whole spectrum of time scales.

\section{Numerical Simulations}

In order to study the influence of the planet's eccentricity on
the spreading of the planetesimals we performed integrations for 
$0.0\leq e_p \leq 0.5$ with $\Delta e_p = 0.1$. 
In Figures 1 to 6 are presented representative snapshots of the temporal evolution of the semi-major axis and eccentricity of the planetesimals, where $\tau=10^6$. There is a strong connection between the orbital evolution of the planetesimals and
mean motion resonances between the planetesimals and the planet. 
In each plot are indicated the location of the semi-major axis for the main mean motion resonances.

In Figure 1 are presented $a\times e$ diagrams for the simulation with $e_p=0$.
The diagrams show the state of the planetesimals at t=0, at t=$1\times 10^3$yr, at t=$2\times 10^4$yr and at t=$1.5\times 10^5$yr. All planetesimals start with circular orbits and at some stage are captured in a resonance. Then, their eccentricities start to growth until they reach a very high value and be removed from the system by collision or ejection. In this case ($e_p=0$) the planetesimals are captured in one of the following resonances: 3:2, 5:3 and 2:1. The 2:1 resonances is the one that plays the main role in the orbital evolution of the planetesimals.

In Figures 2 to 4 are presented $a\times e$ diagrams for the simulations with $e_p=0.1$, 0.3, and 0.5, respectively. 
As the value of the planet's eccentricity increases more mean motion resonances, located further from the planet, become important for the dynamics of the planetesimals. While those closer to the planet become less important.
In the case of $e_p=0.1$ the resonances 5:2 and 3:1 are also important, but the resonance 2:1 is still the one that plays the main role in the orbital evolution of the planetesimals. For $e_p=0.2$ the resonance 4:1 is also important while the resonances 3:2 and 5:3 contribute very little in the orbital evolution of the planetesimals.
For $e_p=0.3$, 0.4 and 0.5 other resonances become also important (7:2, 5:1, 6:1, ...). 
Then the orbital evolution of the disc of planetesimals is affected by several resonances and is not dominated by a particular one. In these cases the planetesimals of the outer part of the disc (a>0.65AU) are spread without being captured in resonance.

\begin{figure*}
\begin{center}
\includegraphics[width=8.5cm]{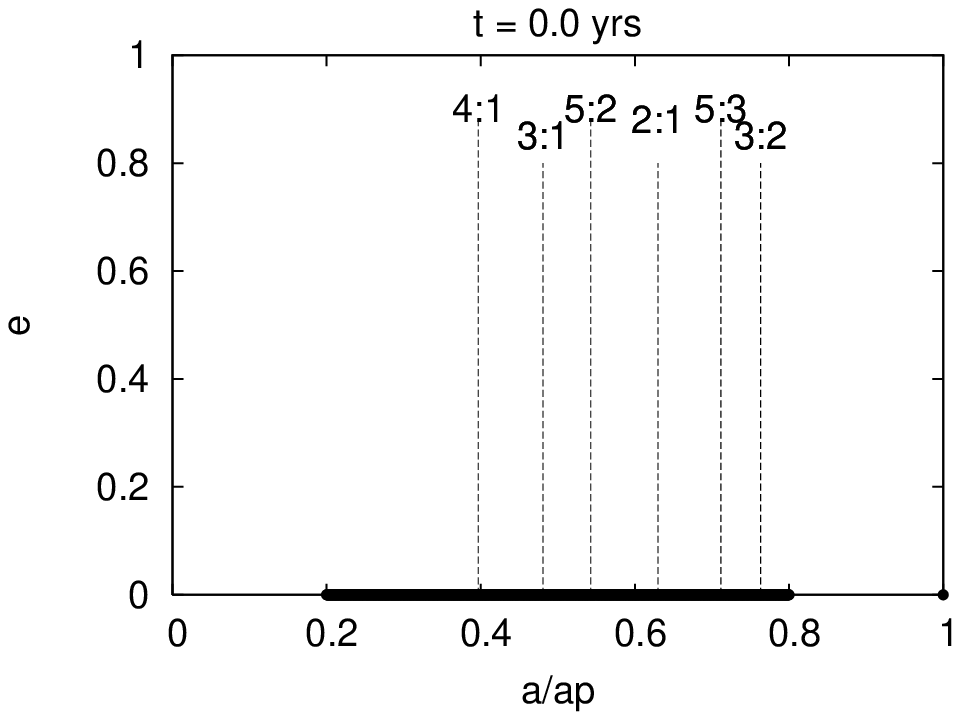}
\includegraphics[width=8.5cm]{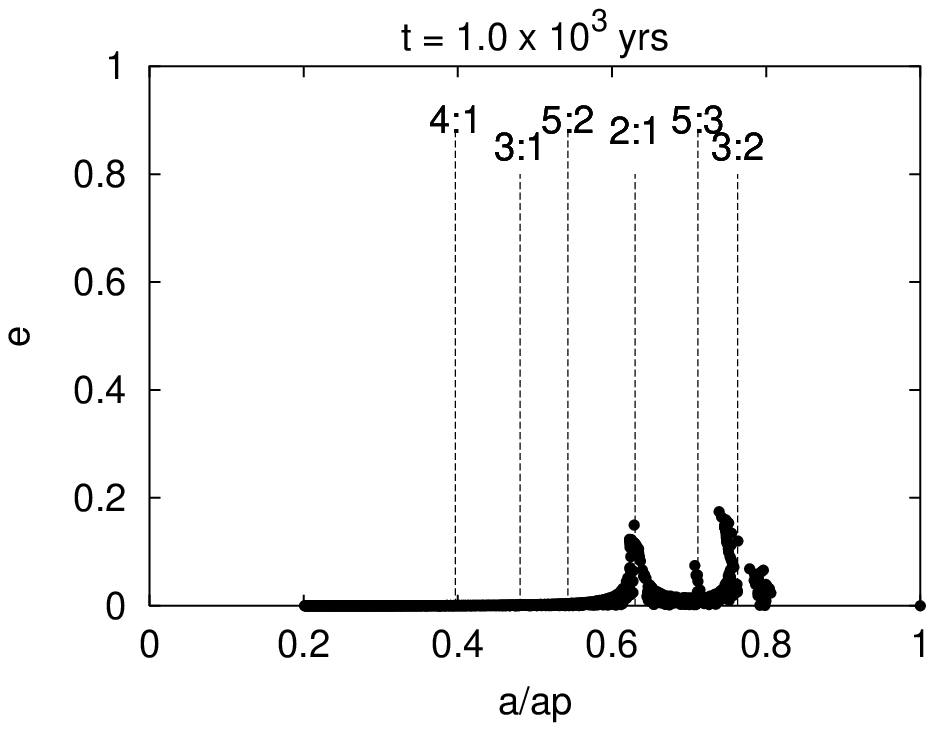}
\includegraphics[width=8.5cm]{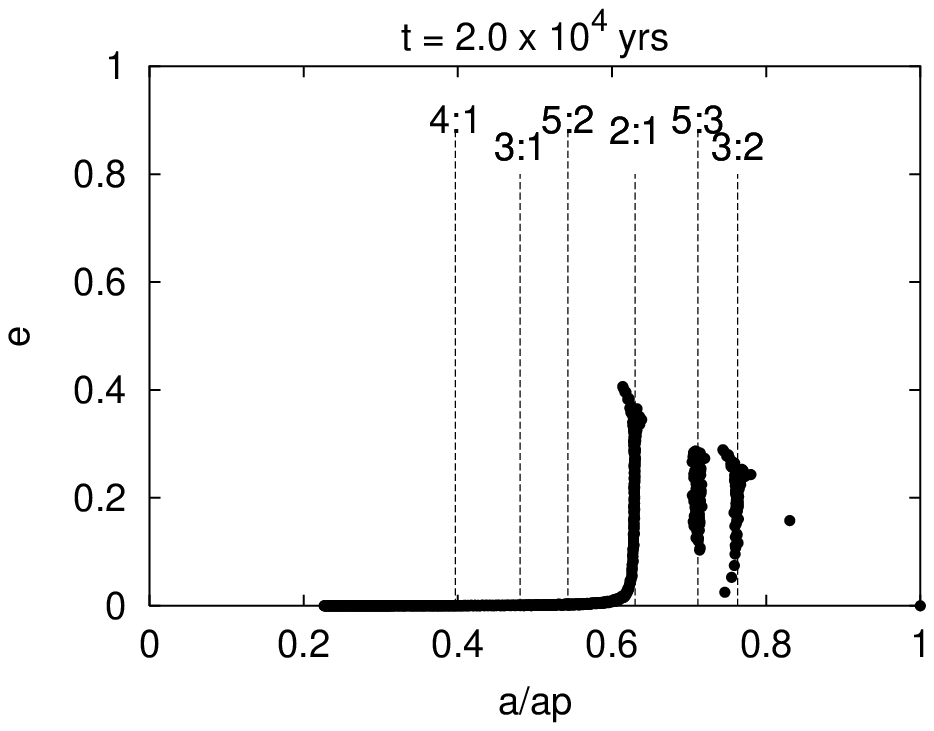}
\includegraphics[width=8.5cm]{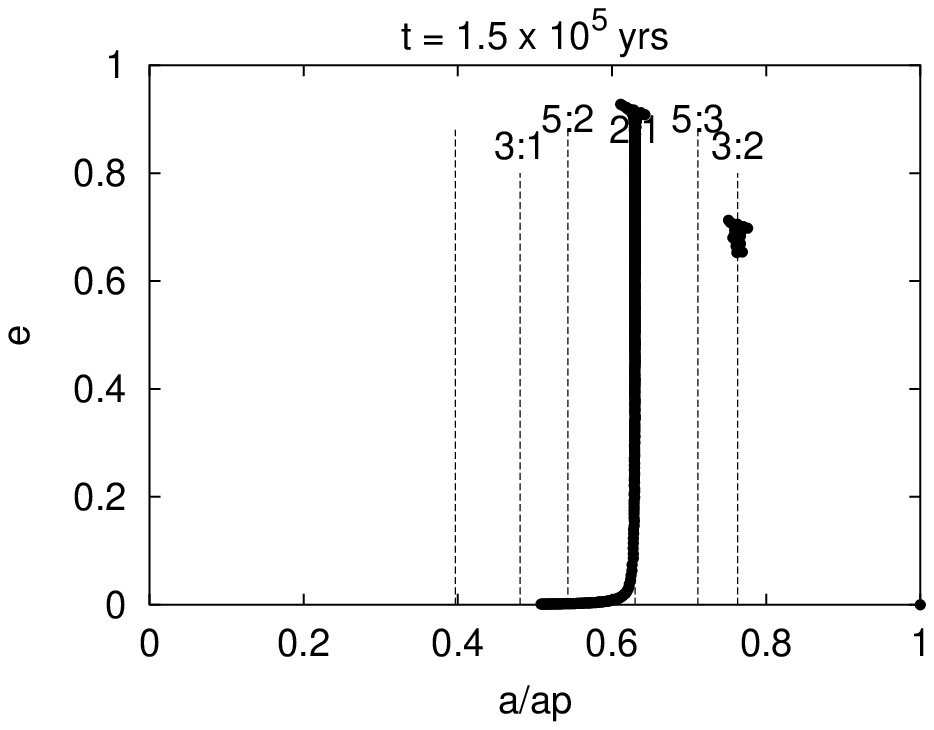}
\end{center}
\caption{\label{figura1}: Temporal evolution of the semi-major axis and eccentricity of the planetesimals for the simulation with $m_p=M_{\rm Jupiter}$, $\tau=10^6$yr and  $e_p=0$.
The $a\times e$ diagrams show the state of the planetesimals at t=0, at t=$1\times 10^3$yr, at t=$2\times 10^4$yr and at t=$1.5\times 10^5$yr. 
In each plot are indicated the location of the semi-major axis for the main mean motion resonances.}
\end{figure*}

\begin{figure*}
\begin{center}
\includegraphics[width=8.5cm]{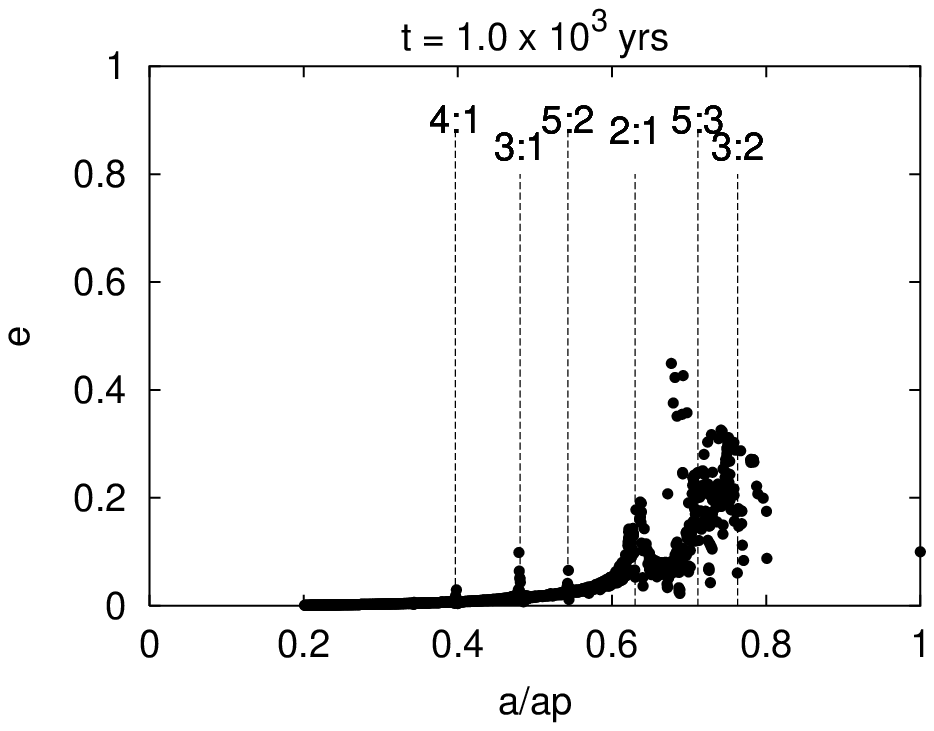}
\includegraphics[width=8.5cm]{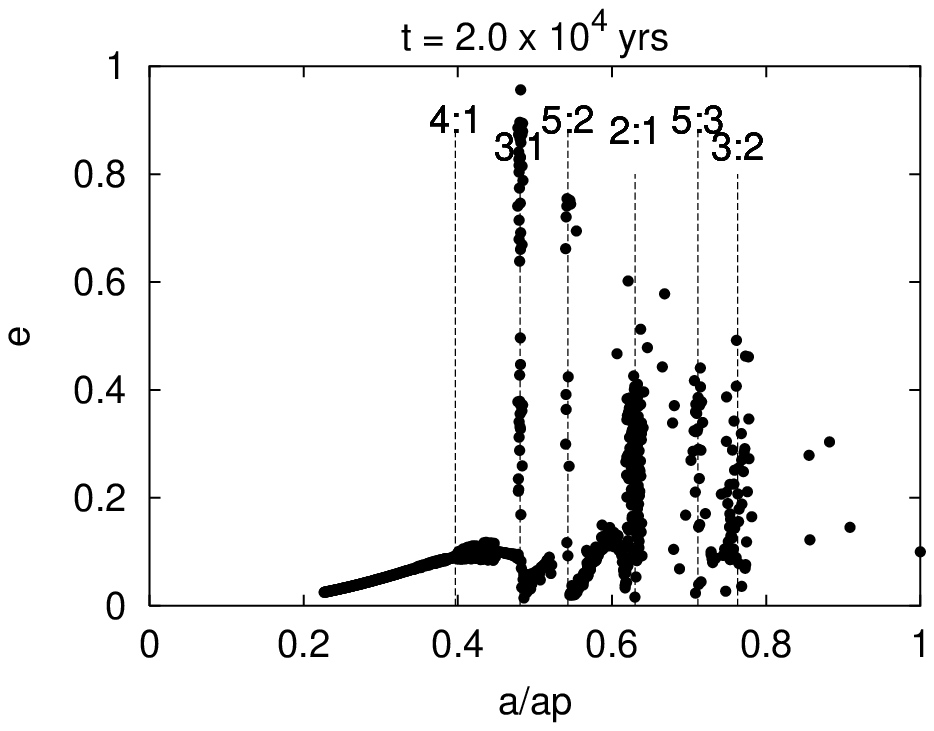}
\includegraphics[width=8.5cm]{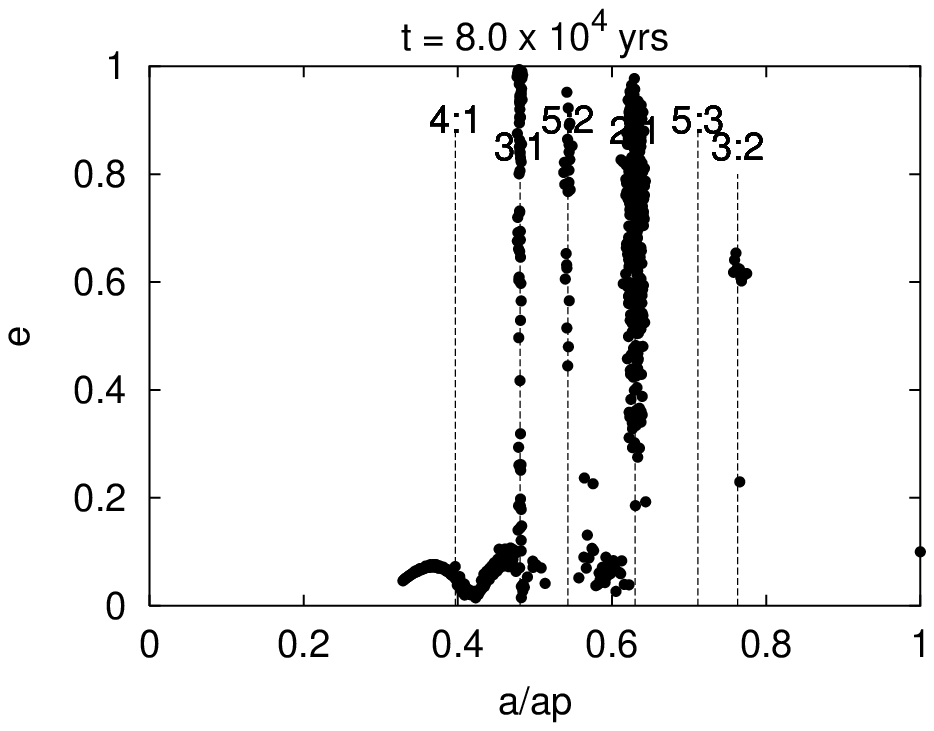}
\includegraphics[width=8.5cm]{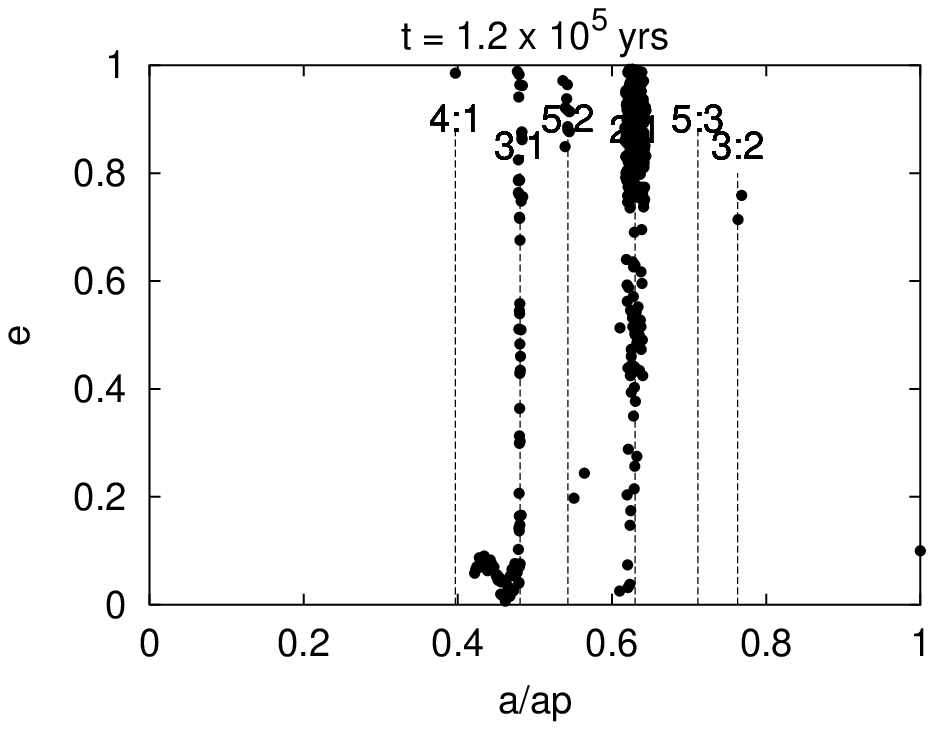}
\end{center}
\caption{\label{figura2}: Temporal evolution of the semi-major axis and eccentricity of the planetesimals for the simulation with $m_p=M_{\rm Jupiter}$, $\tau=10^6$yr and $e_p=0.1$.
The $a\times e$ diagrams show the state of the planetesimals at t=$1\times 10^3$yr, at t=$2\times 10^4$yr, at t=$8\times 10^4$yr and at t=$1.2\times 10^5$yr. 
In each plot are indicated the location of the semi-major axis for the main mean motion resonances.}
\end{figure*}

\begin{figure*}
\begin{center}
\includegraphics[width=8.5cm]{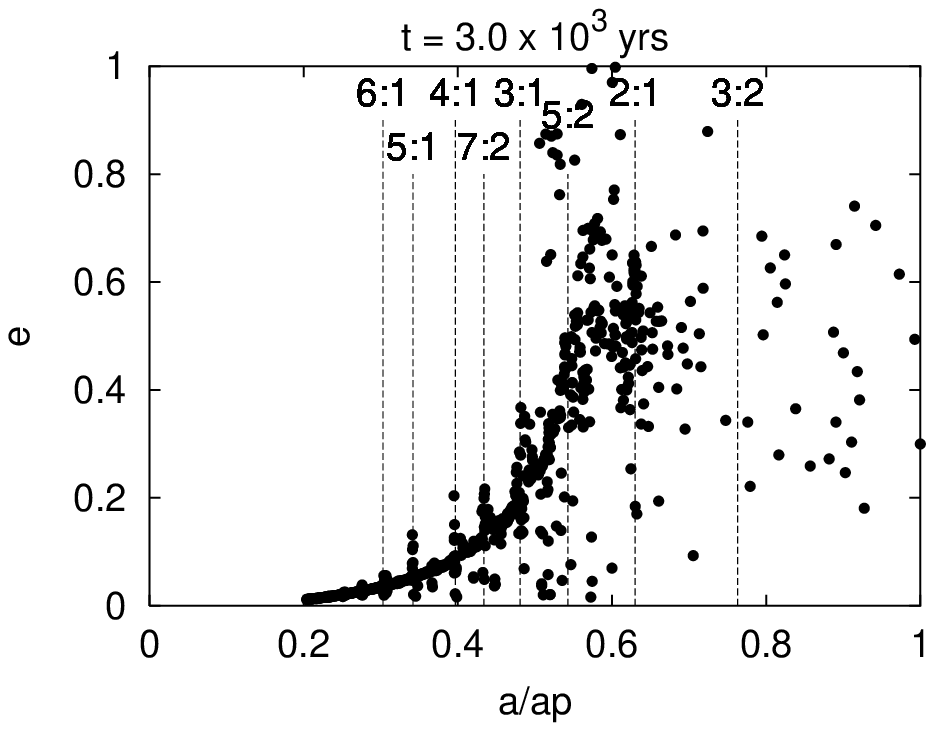}
\includegraphics[width=8.5cm]{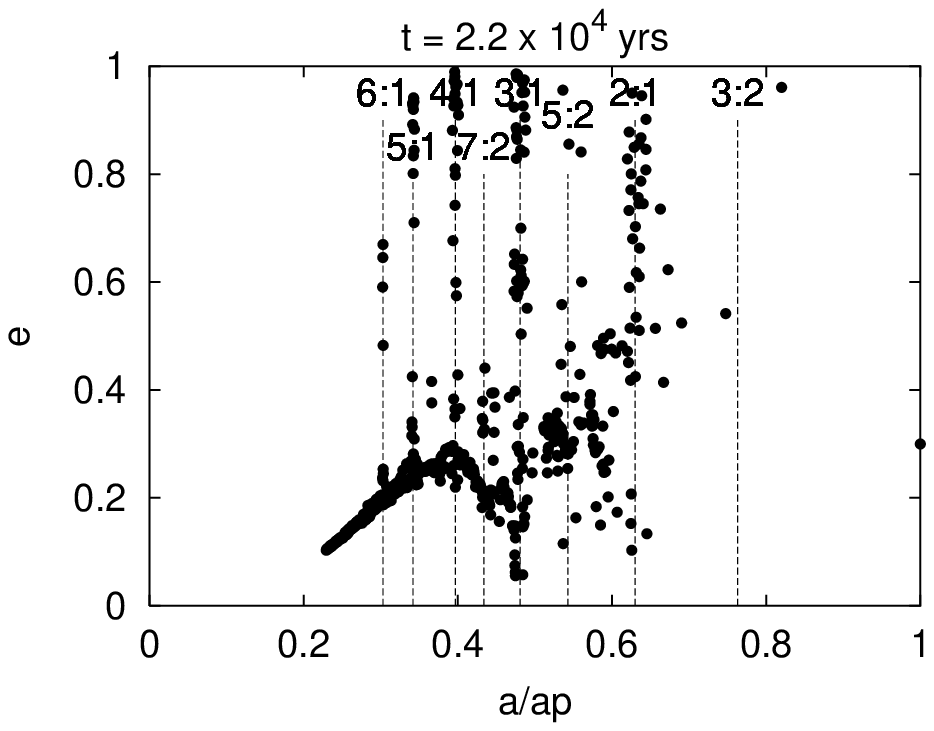}
\includegraphics[width=8.5cm]{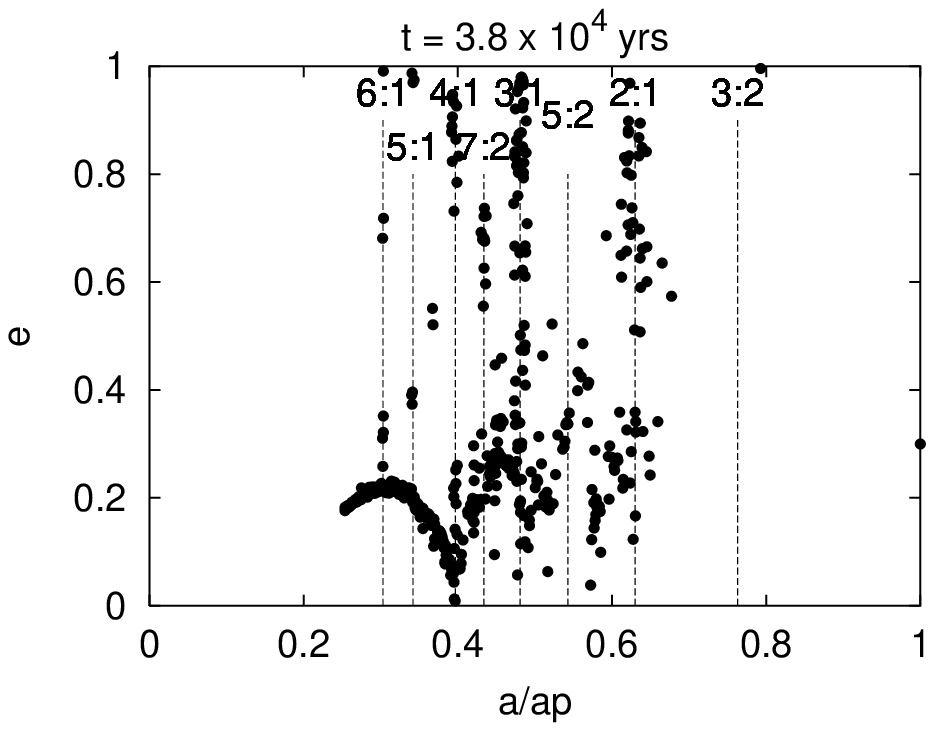}
\includegraphics[width=8.5cm]{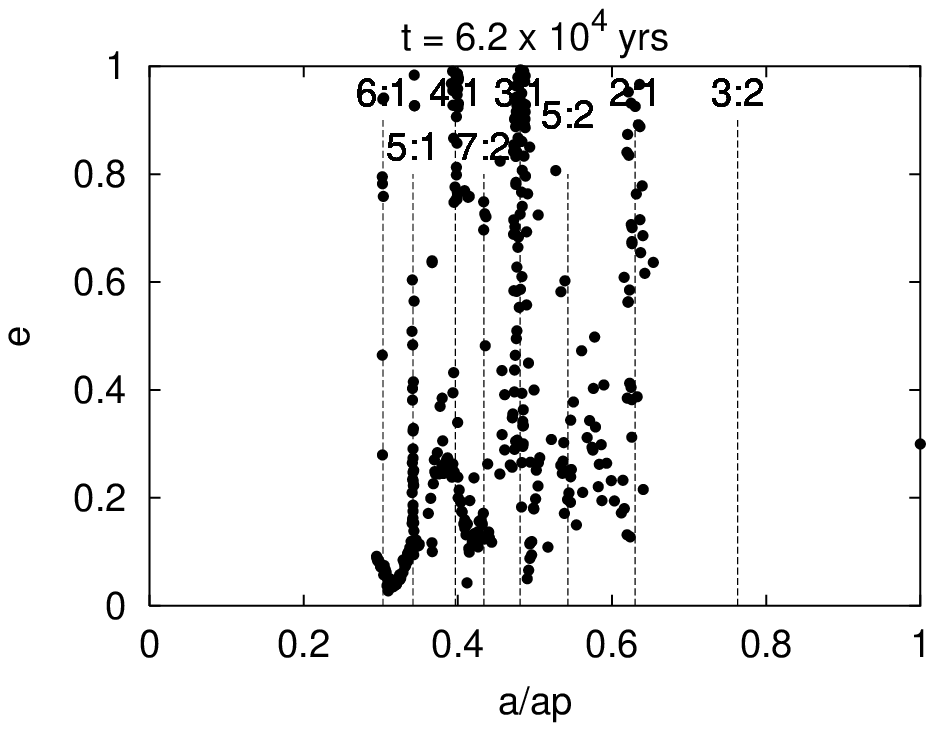}
\end{center}
\caption{\label{figura4}: Temporal evolution of the semi-major axis and eccentricity of the planetesimals for the simulation with $m_p=M_{\rm Jupiter}$, $\tau=10^6$yr and $e_p=0.3$.
The $a\times e$ diagrams show the state of the planetesimals at t=$3\times 10^3$yr, at t=$2.2\times 10^4$yr, at t=$3.8\times 10^4$yr and at t=$6.2\times 10^4$yr. 
In each plot are indicated the location of the semi-major axis for the main mean motion resonances.}
\end{figure*}

\begin{figure*}
\begin{center}
\includegraphics[width=8.5cm]{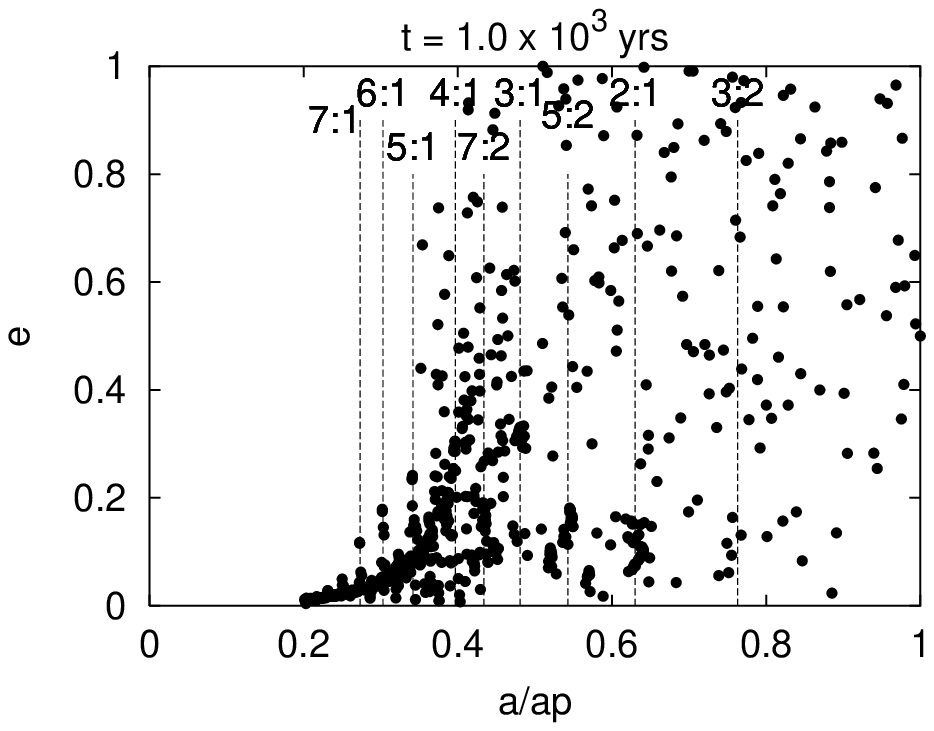}
\includegraphics[width=8.5cm]{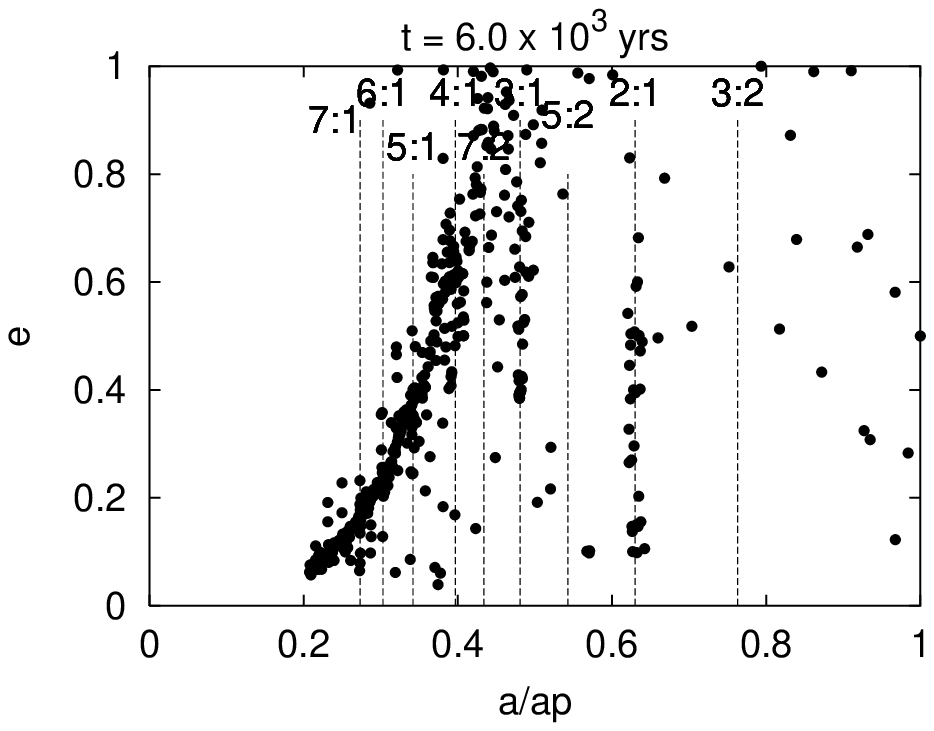}
\includegraphics[width=8.5cm]{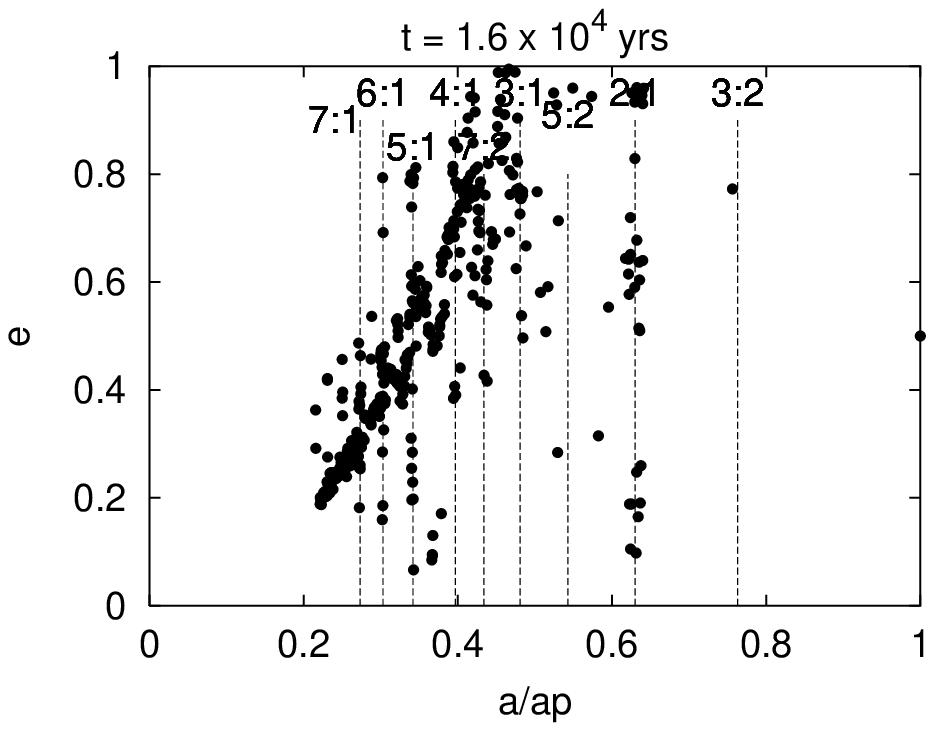}
\includegraphics[width=8.5cm]{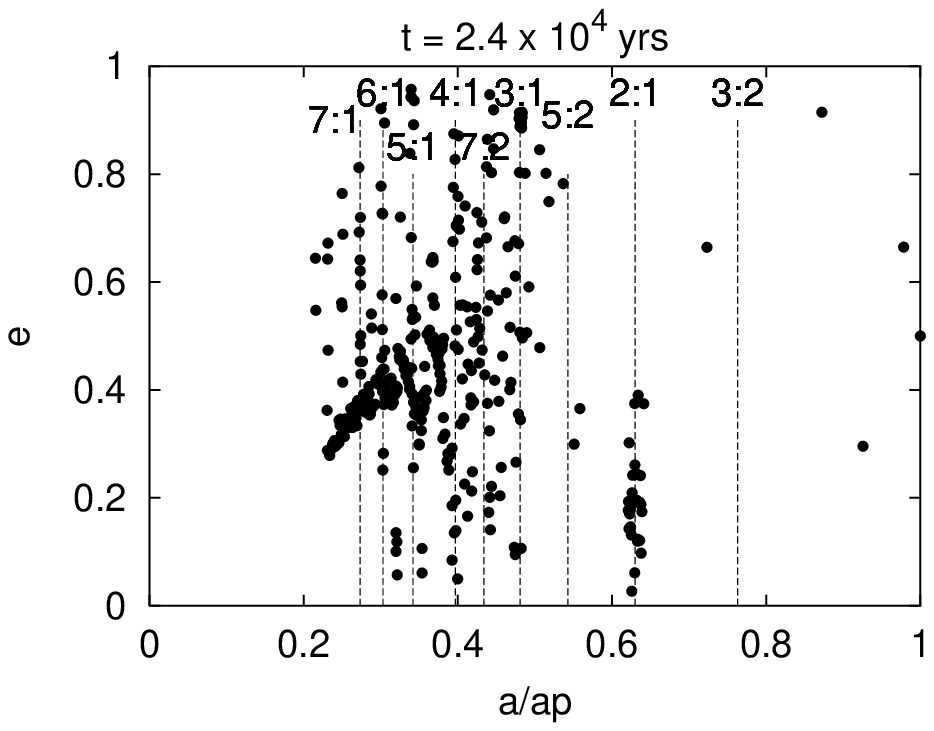}
\end{center}
\caption{\label{figura6}: Temporal evolution of the semi-major axis and eccentricity of the planetesimals for the simulation with $m_p=M_{\rm Jupiter}$, $\tau=10^6$yr and $e_p=0.5$.
The $a\times e$ diagrams show the state of the planetesimals at t=$1\times 10^3$yr, at t=$6\times 10^3$yr, at t=$1.6\times 10^4$yr and at t=$2.4\times 10^4$yr. 
In each plot are indicated the location of the semi-major axis for the main mean motion resonances.}
\end{figure*}

Along each numerical simulation the number of planetesimals goes decreasing with time. The temporal evolution of the remaining percentage of planetesimals in the system is shown in Figure 5. Each plot of Figure 5 corresponds to the simulation with a given $e_p$, where $m_p=M_{\rm Jupiter}$ and $\tau=10^6$yr. The plots show that with the increase of the planet's eccentricity the number of remaining planetesimals in the system decreases faster.

\begin{figure*}
\begin{center}
\includegraphics[width=8.5cm]{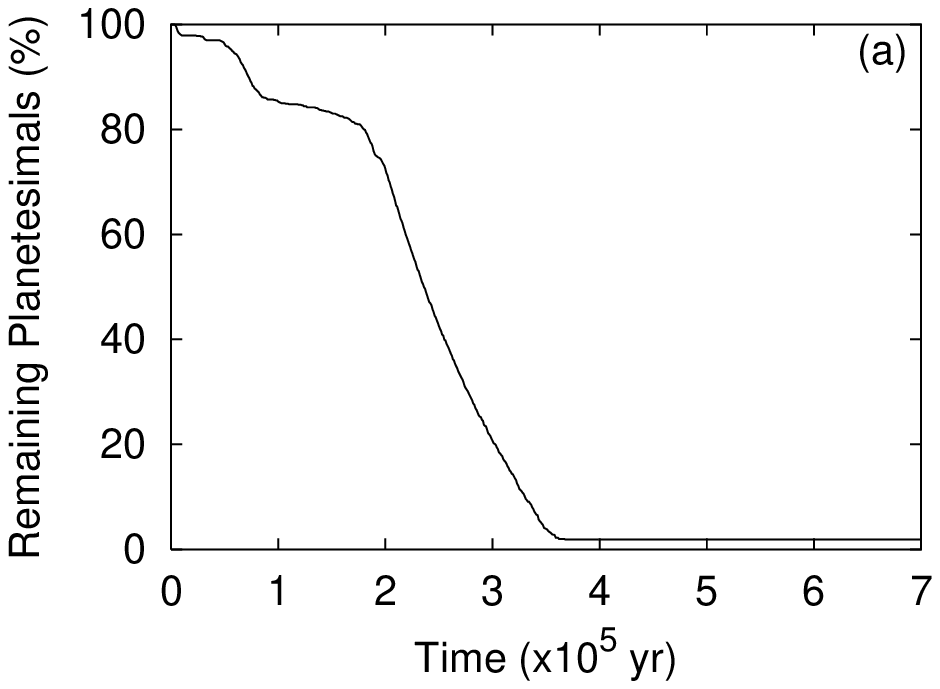}
\includegraphics[width=8.5cm]{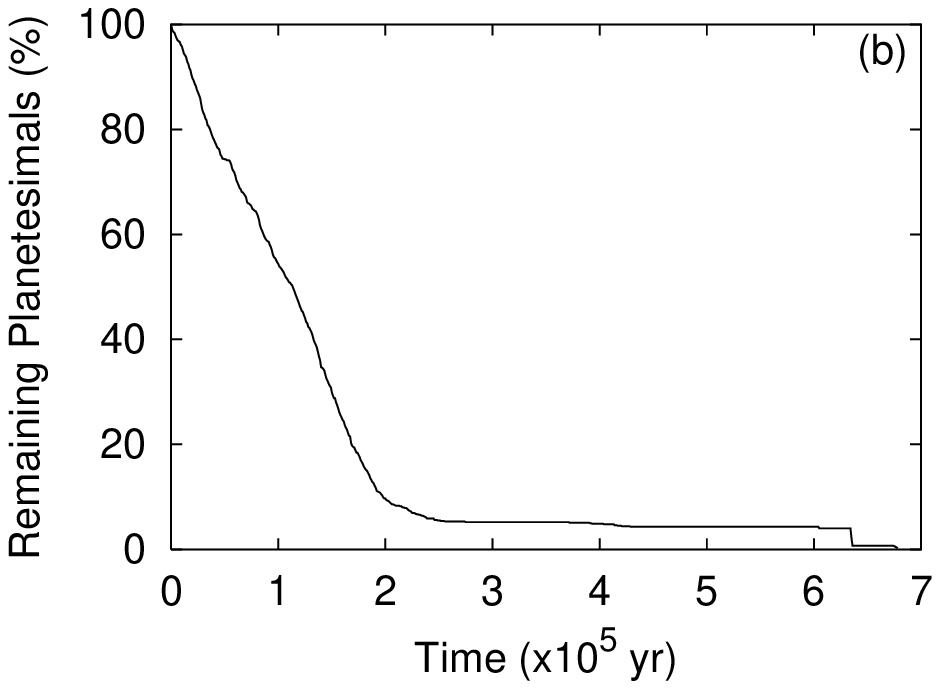}
\includegraphics[width=8.5cm]{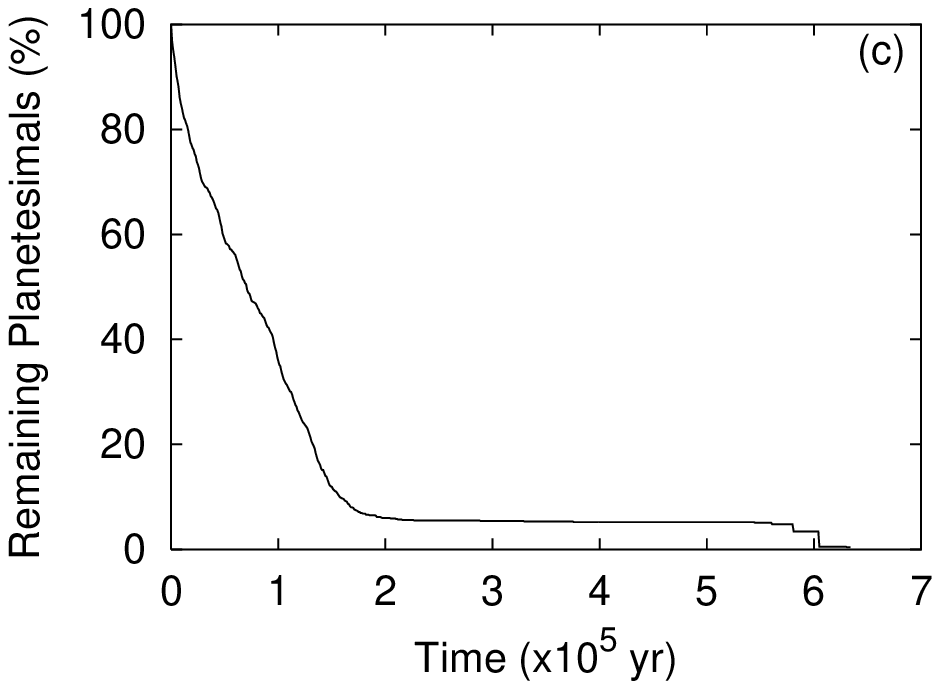}
\includegraphics[width=8.5cm]{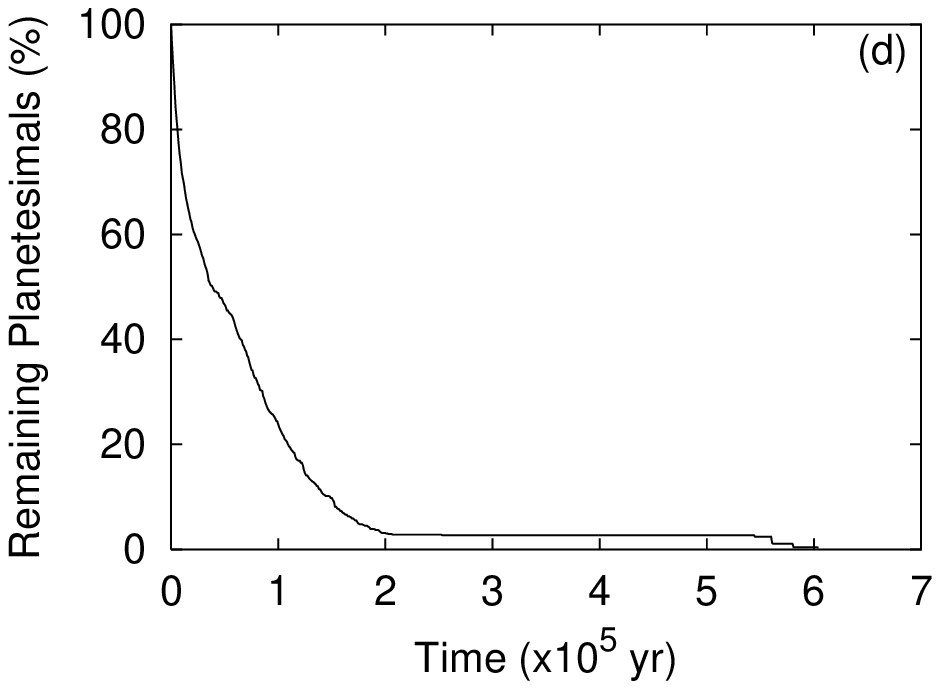}
\includegraphics[width=8.5cm]{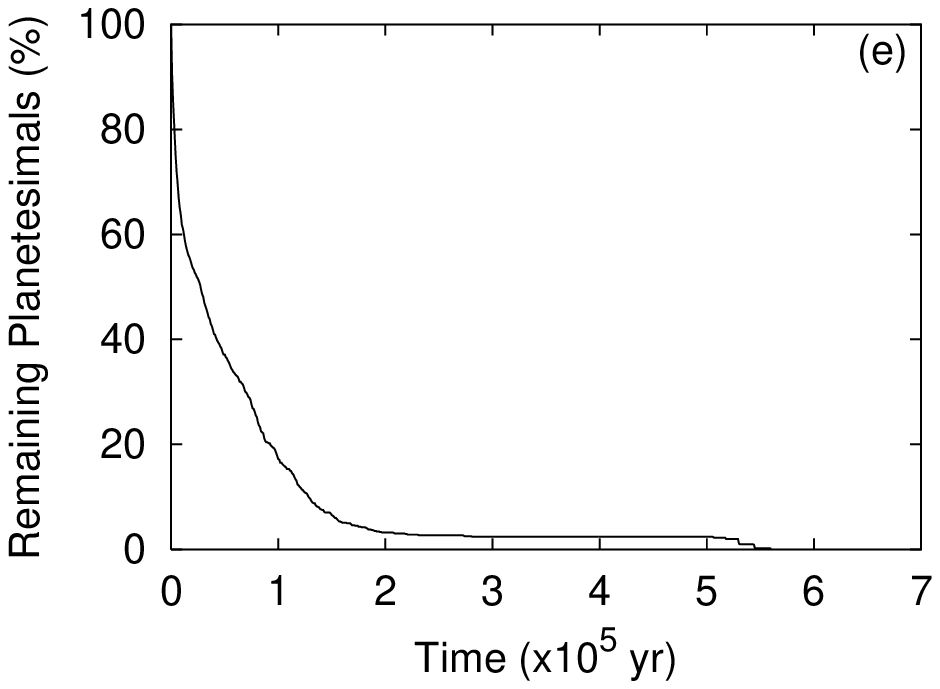}
\includegraphics[width=8.5cm]{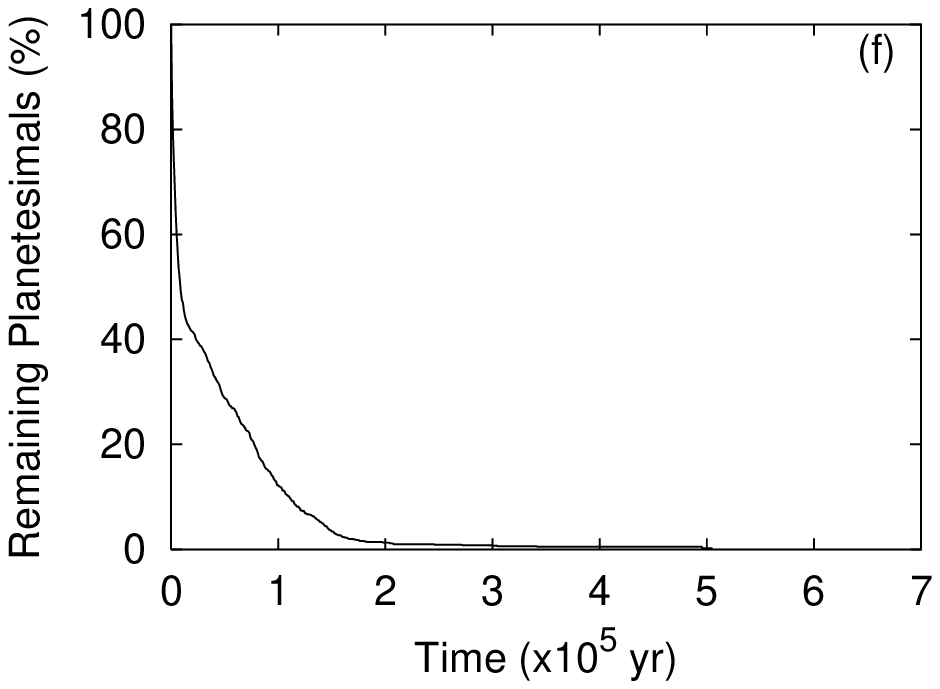}
\end{center}
\caption{\label{figura7}: Temporal evolution of the remaining planetesimals in the system. These are the results from the simulations also presented in Figures 1 to 4, where $m_p= M_{\rm Jupiter}$ and $\tau=10^6$yr . The eccentricity of the planet is: a) $e_p = 0$, b) $e_p = 0.1$, c) $e_p = 0.2$, d) $e_p = 0.3$, e) $e_p = 0.4$ and f) $e_p = 0.5$.}
\end{figure*}

In order to study the influence of the migration speed on
the spreading of the planetesimals we performed integrations 
for different values of timescales,
from $\tau=10^2$ yr to $\tau=10^6$ yr with $\Delta\tau=10$ yr.
Simulations for $\tau=10^7$ yr are computationally too expensive.
We made a few simulations with $\tau=10^7$ yr and the results are not much 
different from those for $\tau=10^6$ yr.

In Figure 6 we present the statistics of the results in terms of percentage of collisions 
with the star, with the planet and ejections from the system.
For slow migrations ($\tau=10^5$yr and $\tau=10^6$yr) the percentage of collisions with the planet decrease with the increase of the planet's eccentricity.
For $e_p=0$ and $e_p=0.1$ most of the planetesimals were captured in the 2:1 resonance and more than 65\% of them collided with the star.
For very fast migrations ($\tau=10^2$yr and $\tau=10^3$yr) 
there is no capture in mean motion resonances, independently of the value of $e_p$.
Then, due to the planet's migration
the planetesimals suffer close approaches with the planet and 
more than 80\% of them are ejected from the system.

Quillen (2006) developed analytical predictions of capture in resonances. His results show that raising $e_p$ above a certain limit, $e_{lim}$, may prevent the resonance capture. He also predicts that the capture probability is reduced for faster migration. Our results corroborate his predictions. The capture probability for resonance also depends on the planets mass (Quillen, 2006), which will be changed in the next section.

\begin{figure*}
\begin{center}
\includegraphics[width=8.5cm]{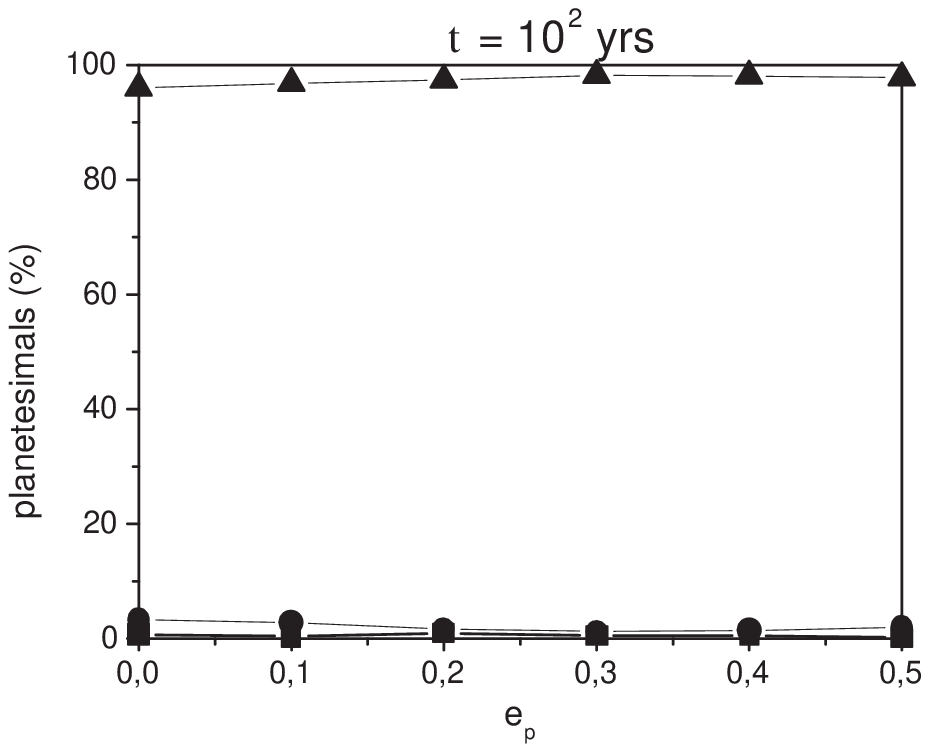}
\includegraphics[width=8.5cm]{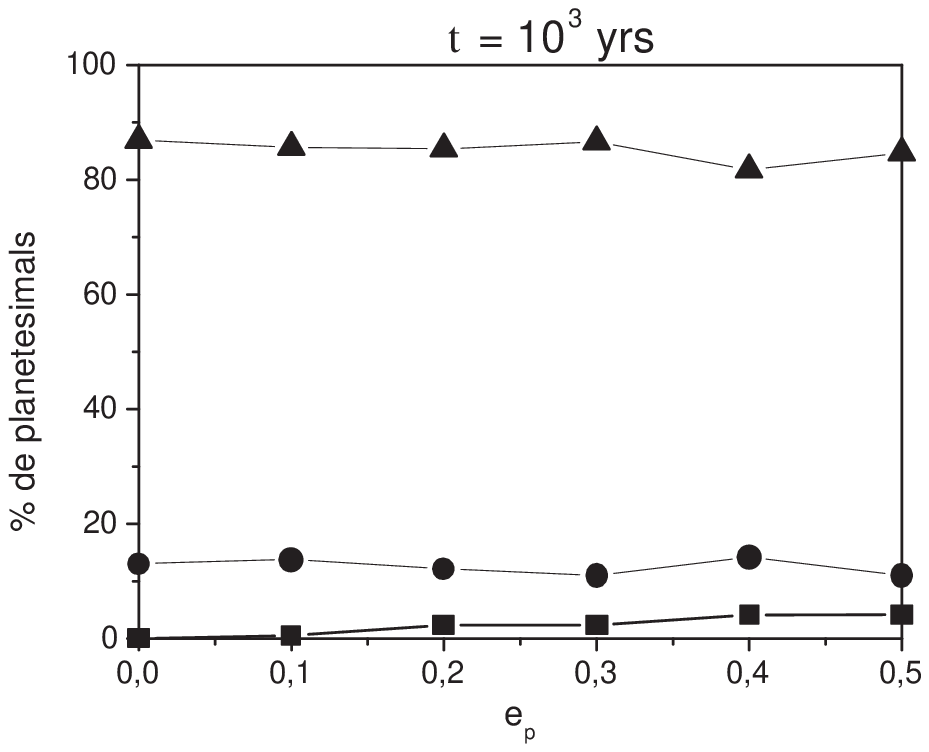}
\includegraphics[width=8.5cm]{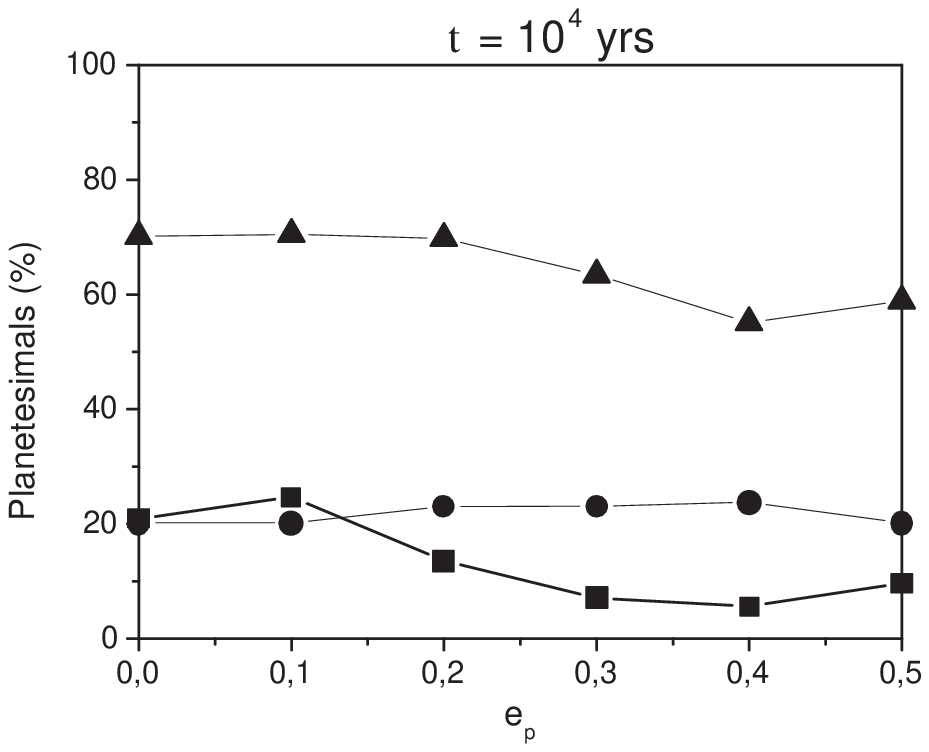}
\includegraphics[width=8.5cm]{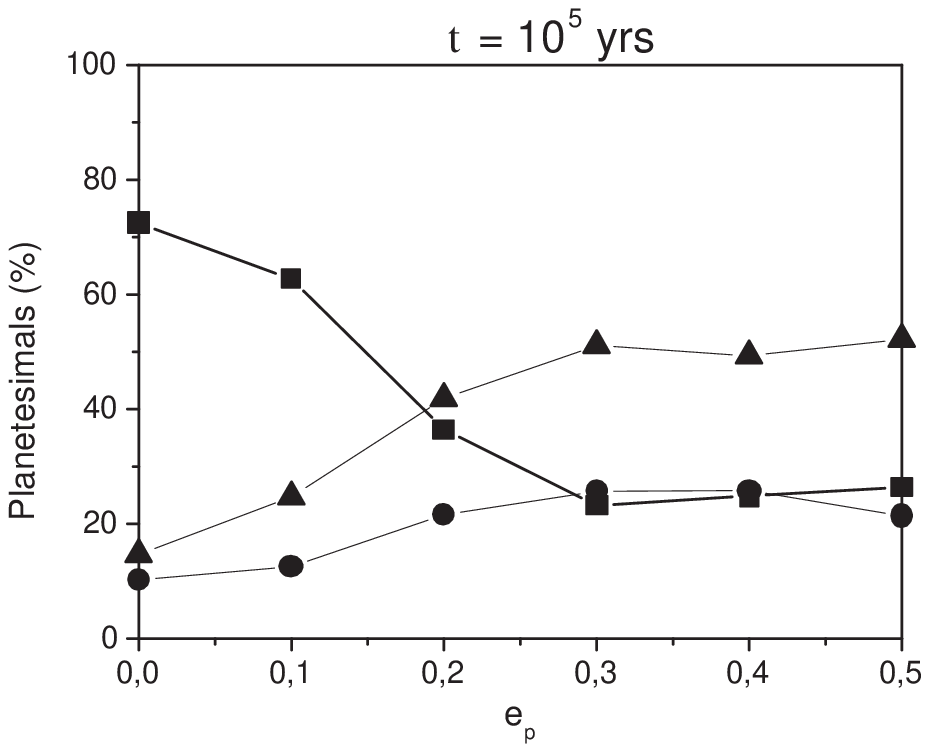}
\includegraphics[width=8.5cm]{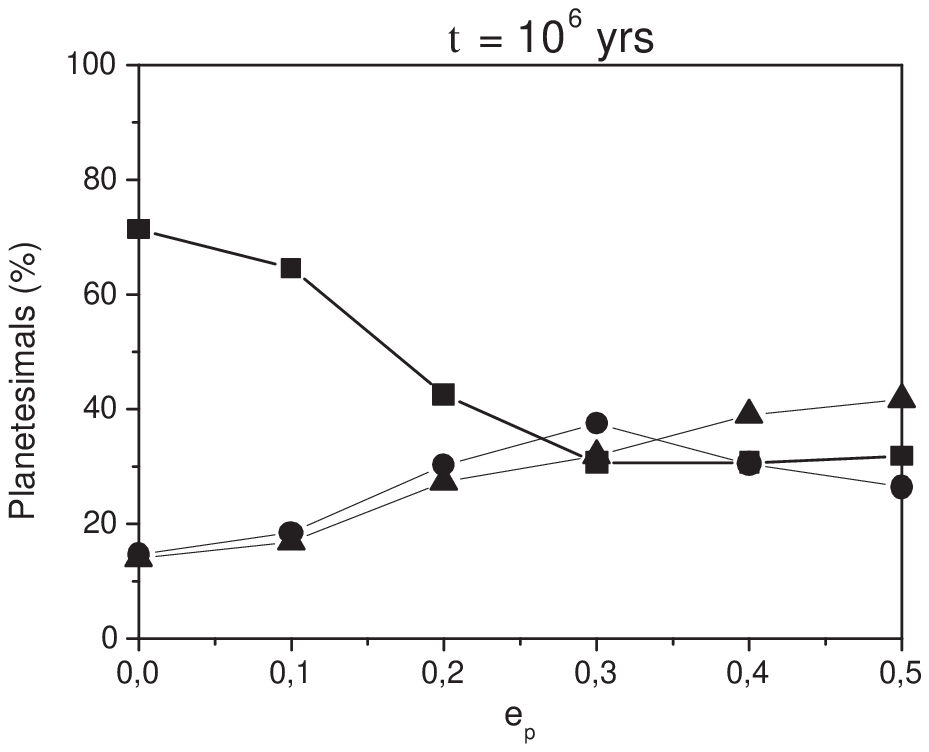}
\includegraphics[width=8.5cm]{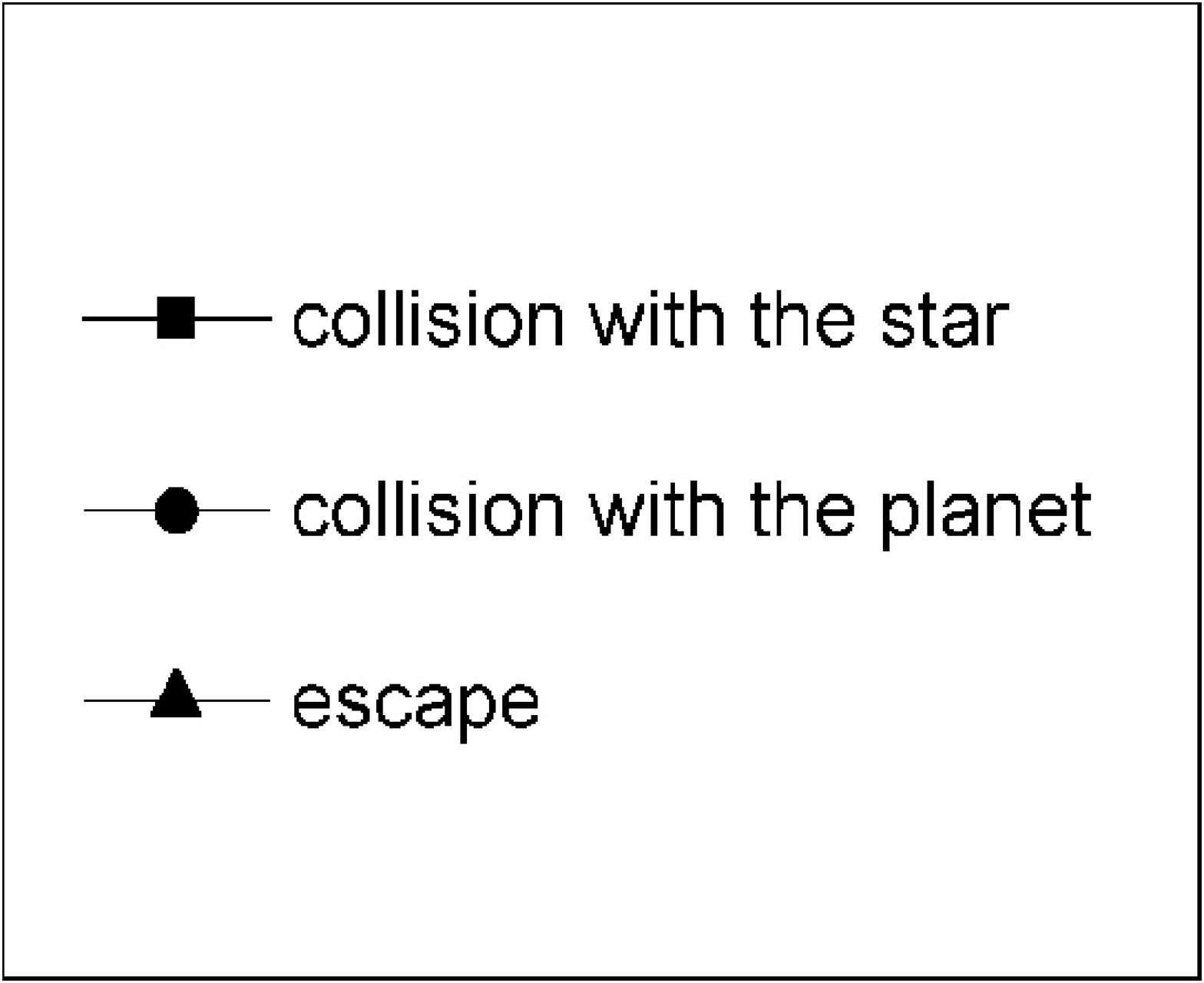}
\end{center}
\caption{\label{figura8}: Planetesimals end state for the simulations with $m_p=M_{\rm Jupiter}$. In each plot is given the percentage of planetesimals that collided with the star (triangle), that collided with the planet (circle) and that escaped (square). The values are given as a function of the planet's eccentricity. On the top of each plot is shown the corresponding time scale for the migration of the planet.}
\end{figure*}

\begin{table*}
  \centering
  \caption{\textbf{Summary of the results for the simulations with $m_p=M_{\rm Jupiter}$ and $e_p=0$, its is given the percentage 
of collisions of planetesimals with the star as a function of the planet's migration 
time scale. It is also shown comments on the problems related with each case.
}}\label{Tabela1}\vskip0.5cm
\begin{tabular}{|l|c|c|c|c|c|c|}\hline
  $\tau$(yr) & Migration Model & Star Collision (\%)  & Problem \\ \hline\hline
  $10^2$ & Gas           & >5 & low star collision rate \\
  $10^3$ & Gas           & >5 & low star collision rate \\
  $10^4$ & ---           & ~20 & ---                     \\
  $10^5$ & Planetesimals & ~70 & disc too massive  \\
  $10^6$ & Planetesimals & ~70 & disc too massive  \\
\hline\hline
\end{tabular}
\end{table*}

\section{Mass Problems}

The minimum mass in scattering planetesimals necessary to produce the
migration of a planet can be given by (Adams \& Laughlin 2003)
$$M = m_p \ln(a_0/a_f)\eqno$$
where $m_p$ is the mass of the planet, $a_0$ the initial semi-major axis of
the planet and $a_f$ its final semi-major axis. From our simulations ,
$a_0= 5$ AU, $a_f$
= 0.01 AU and $m_p$ equal to one Jupiter we obtain that $M$ is equal to about 6
$M_{\rm Jupiter}$. As we will see after the migration of a Jupiter mass planet
up to the
pericentre distance of 0.01 AU (2.15 solar radius) demands extremely
large disc masses. However, this is not the case for a 0.1 $M_{\rm Jupiter}$
migrating planet. Davis (2005) proposed a cumulative mass growth in
the primitive solar nebula model such that the total mass of the disc
up to Jupiters position would be about 3838 $M_{\rm Earth}$. Adopting a
canonical, gas to dust ratio of ~100, we expect only 1\% in the mass of
planetesimals. Then the mass of planetesimals in our adopted disc
model from Davis (2005) would be 38 $M_{\rm Earth}$. Our simulations show for a
planet with Jupiters mass, in near circular orbit, migration from 5
AU to 0.01 AU in a time scale of $\tau > 10^5$ yr, that about ~80\% of
the planetesimals collide with the star. However, the minimum mass
required for a Jupiter mass planet to migrate up to that pericentre
distance is, as we see above, of the order of 6 $M_{\rm Jupiter}$. This
represent for ( $M_{\rm Jupiter}$= 318 $M_{\rm Earth}$) a disc mass of 1908 $M_{\rm Earth}$ of
planetary mass, which represent 50 times our adopted primitive solar
nebula. This is unrealistic. The situation is much better for a
migrating planet of 0.1 $M_{\rm Jupiter}$ with a required mass of 190 $M_{\rm Earth}$.

As an example, lets consider which will be the requirements for the maximum
concentration of observed hot jupiters exoplanets at ~0.05 AU.
In this case, The necessary minimum mass is 1464 $M_{\rm Earth}$ which is a factor 38
larger than our primitive solar adopted mass and represent a disc of
0.4 solar mass, also unrealistic. For the case of 0.1 $M_{\rm Jupiter}$ planet,
for the same migration distance requires a disc (up to Jupiter
distance) of 0.04 solar mass which could be appropriate.

Considering this, we search for possibilities for a slow migration
( $\tau$ = $10^5$ - $10^6$ yr) of a 0.1 $M_{\rm Jupiter}$ mass planet. The simulations
results for this case are shown in Table 2. Here, the planet is set to
migrate from 5 AU up to 0.01 AU in principle with $\tau$= $10^6$ yr. We
performed simulations  for $e_p = 0$ to $e_p = 0.5$ with $\Delta
e_p = 0.1$. In columns 5 and 6 are given the time and the value of
the semi-major axis of the planet when the last planetesimal is
removed from the system. For $e_p = 0.1$ (Figure 7) the last planetesimal is
removed at $t = 1.71\times 10^5$ yr, when the planet has  $a_f = 1.72$ AU. In
order to produce such migration the mass of ejected planetesimals
(Equation 1) has to be of the order of 34 $M_{\rm Earth}$. We found that in
this simulation 50\% of the planetesimals collide with the star and
about 45\% are ejected (Table 2). Therefore, in such case a disc of
planetesimals with mass of about 70 $M_{\rm Earth}$ would not be too massive
for the present models (Davis 2005) and would produce the planets
migration required to generate the metallicity expected.

\begin{figure*}
\begin{center}
\includegraphics[width=8.5cm]{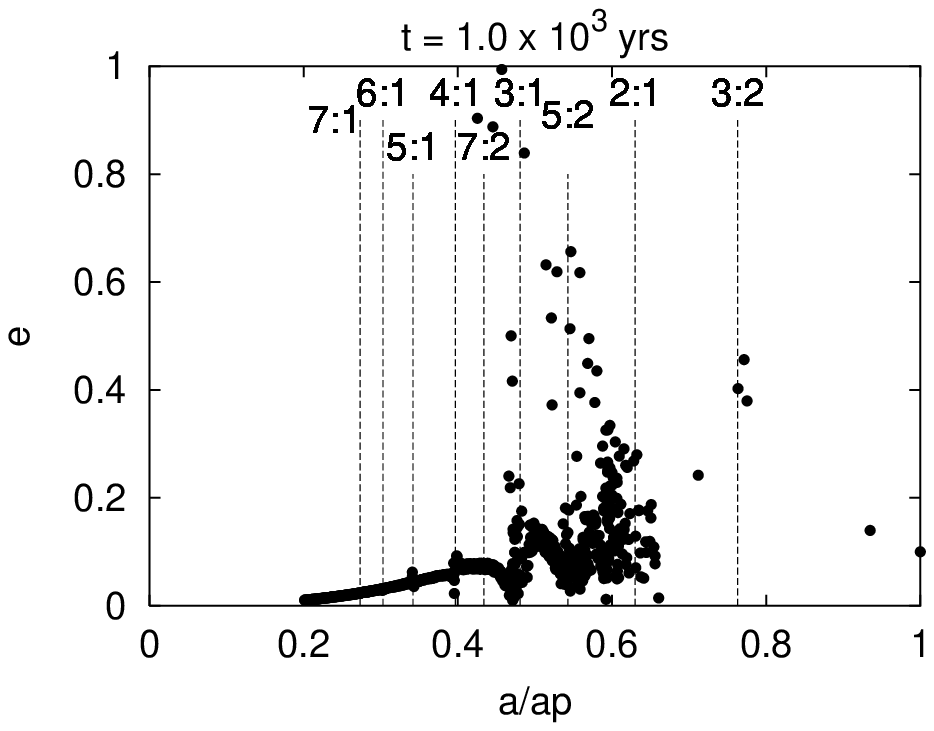}
\includegraphics[width=8.5cm]{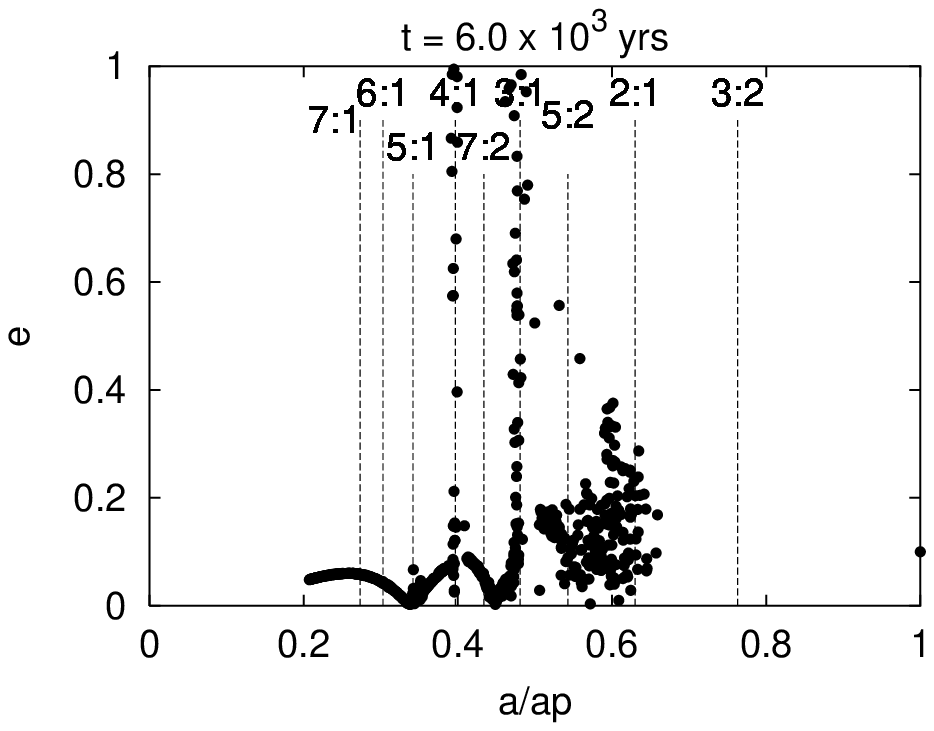}
\includegraphics[width=8.5cm]{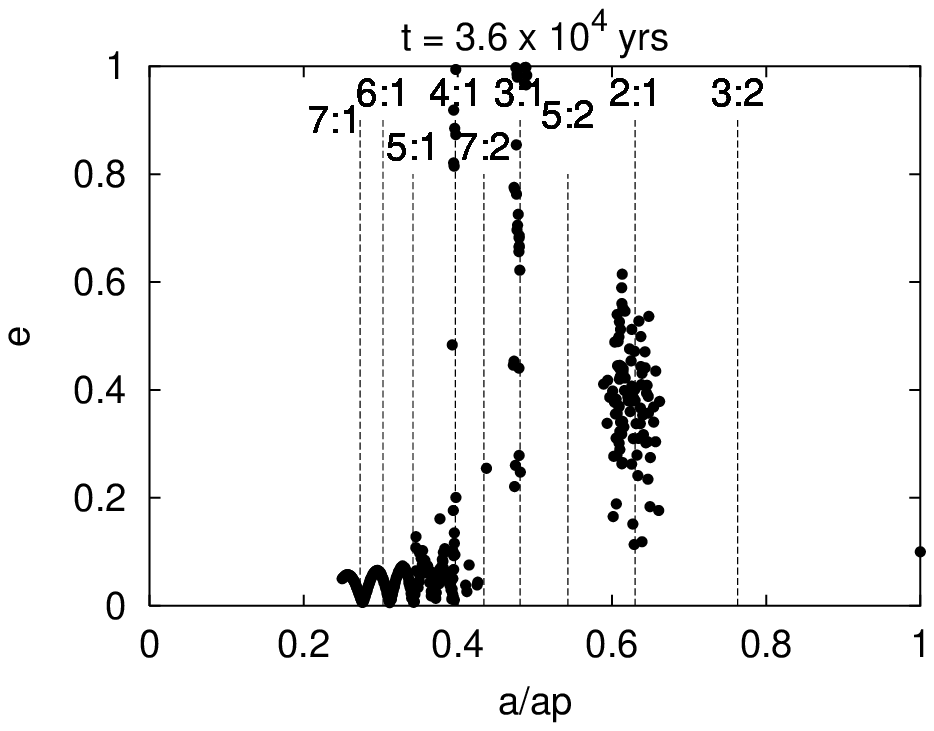}
\includegraphics[width=8.5cm]{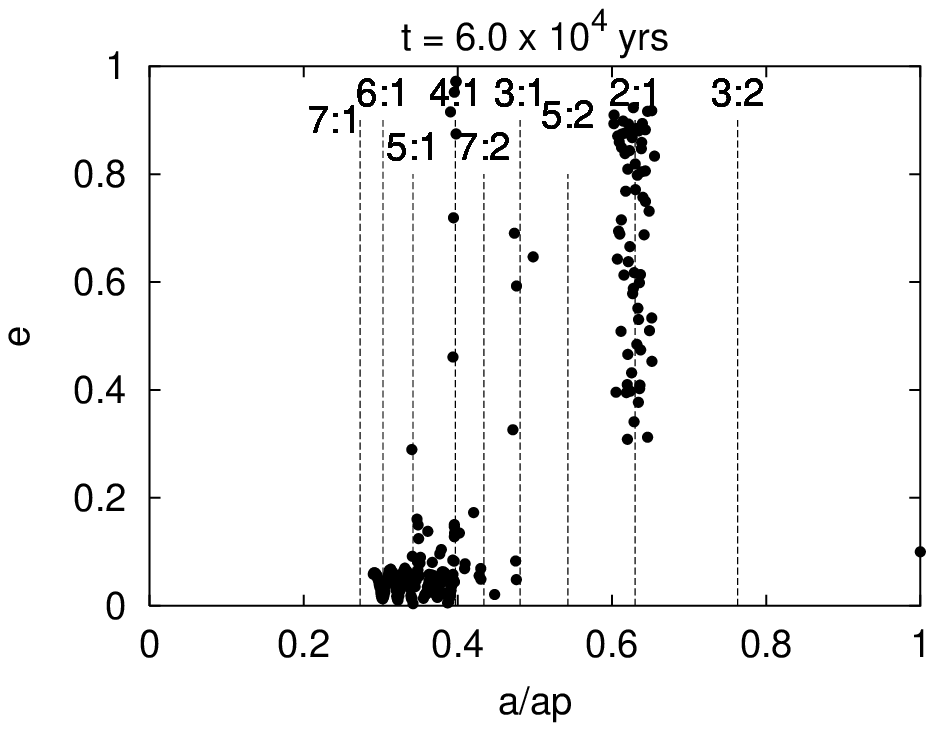}
\end{center}
\caption{\label{figura9}: Temporal evolution of the semi-major axis and eccentricity of the planetesimals for the simulation with $m_p=0.1M_{\rm Jupiter}$ , $\tau=10^6$yr and $e_p=0.1$.
The $a\times e$ diagrams show the state of the planetesimals at t=$1\times 10^3$yr, at t=$6\times 10^3$yr, at t=$3.6\times 10^4$yr and at t=$6\times 10^4$yr. 
In each plot are indicated the location of the semi-major axis for the main mean motion resonances.}
\end{figure*}

\section{Metal Enhancement in PTTS}

Any model using injection of rocky planetesimals as a way to enhance stellar
metallicities of young stars, requires a compromise between the epochs of
existence of planetesimals, the time of planet migration (at least for a case
of a single planet as is the case here) and the epoch of diminishing of the
surface convective layer. This last condition is specially important in order
that the metal enhancement be efficient. In fact, very young stars as T Tauri,
are characterized by their large convection zones and these must be stabilized
to their minimum configuration. In that case, the new injected material is
trapped and not diluted as is the case in stars with large convective zones.

Due to a very rapid decrease of convective envelopes of solar type stars, this
mentioned compromise appears between 20 and 30 Myr. This "window of
opportunity" as is clearly shown in Figure 5 in Ford, Rasio \& Sills
(1999)      represent the right time to enhance the metal content,
that in principle is maintained and observed today. A possible
signature of this metallic enrichment has been detected in de la Reza
et al. (2004). Here a collection of G and K stars belonging to two
coeval PTTS associations with ages of 20 and 30 Myr showed G stars
with larger Fe abundances in respect to the lower mass (and larger
convection zones) K type stars. In our scenario, as is also the case
of QH00, the metal enrichment can be produced by particles directly
impinging
in the stars or by the grazing planetesimals with very large
eccentricities. In our simulation we obtain , for the low migrating
rates and depending on the approaching distances to the star, $10^3$ up
to $10^5$ close passages.

QH00 considered that if rocky bodies are larger than one kilometer, they will
survive vaporization. other possible destiny of particles is disruption of
strengthless material. Sridhar \& Tremaine (1992) (see also Asphaug \&
Benz 1996)
found that inviscid planetesimals in parabolic orbits will be
disrupted by the planet, or in our case by the star, if their
pericentric distances are less
than  1.69 $R_s(\rho_s/ \rho_{\rm p})^{1/3}$. Here, $R_s$ is the stellar radius, $\rho_s$ the
stellar density and $\rho_{\rm p}$ the density of the planetesimal. After disruption,
the planetesimals became a kind of "dust" maintaining their same densities.

In view of this and as an example, we can explore which will be the disruption
distances for three types of representative stars. An A type star with (two
solar mass, two solar radius), a solar G star, and a K star (half solar mass
and radius). Their respective densities being 0.35, 141 and 5.6 g/cc. The
corresponding disruption distances will be 0.8, 1.31 and 2.1 stellar radius.
Taking into account that in our simulations, we consider that an impact is
defined when a particle is closer than 0.01 AU (2.15 solar radius), we
have that for our A star with 2 solar radius, practically all
particles penetrate the star and "disruption" will occur inside the
star (of course this is not valid because the physical situation inside
the star must be
different). For a solar type star, particles with a pericentric
distances less than 1.3 solar radius will be disrupted. For K stars
particles already  at distances less than 2 stellar radius will be
disrupted. In any case, it is expected that these particles will
contribute to the metal enrichment of the star.

Now, lets consider which will be the possible metal enhancements for the most
realistic case in which a planet of 0.1 $M_{\rm Jupiter}$ is migrating. In Table 2 are
displayed, for different planets eccentricities, the minimum masses
(Col. 7) following relation 1, required to terminate the migration at
the pericentric distances indicated in (Col.6). The impacting masses
are shown in (Col 8). In case of high penetration efficiency of
particles, we obtain that the best cases are represented by planet
eccentricities 0.0, 0.2 and
0.3. Adopting the Fe enrichment calculated curves in Figure 5 by Ford et
al.(1999), we obtain for the impacting masses, [Fe/H] abundances at 20
and 30 Myr, of the order of 0.18 dex for $e_p$ = 0.0 and $e_p$
= 0.2 and about 0.15 dex for $e_p$ = 0.3. Because for our
simulations shown in Table 2, we need planetary maximum masses of the
order of 100 $M_{\rm Earth}$. Reminding that our primitive solar disc contains
about 38 $M_{\rm Earth}$ of solid matter, we conclude that primitive discs to
produce these enrichments must be larger than a factor 3 than the
solar one. Larger factors, nevertheless possible to exist, are
necessary to explain the observed hot Jupiters exoplanets at ~0.05
AU, with masses below one Jupiter mass, even with sightly larger Fe
abundances.

\begin{table*}
  \centering
  \caption{\textbf{Simulations for a small planet ($m_p=0.1M_{\rm Jupiter}$) migrating from
5AU to 0.01AU in a time scale $\tau=10^6$yr.
In columns 5 and 6 are given the time and the value of the semi-major axis of the planet when 
the last planetesimal is removed from the system. In column 7 is given the minimum masses
(Eq. (1)) required to terminate the migration at
the pericentric distances indicated in column 6. The impacting masses
are shown in column 8.
}}\label{Tabela2}\vskip0.5cm
\begin{tabular}{|l|c|c|c|c|c|c|c|c|}\hline
  $e_p$ & Star Collision  (\%)& Planet Collision (\%)  & Ejected (\%) & $t_{\rm last} (\times 10^5)$yr & 
  $a_{p-\rm last}$(AU) & $M$($M_{\rm Earth}$)   & Impact. Mass ($M_{\rm Earth}$)  \\ \hline\hline
  0.0 & 68.5 & 3.1 & 28.4 & 2.52 & 1.04 & 50 & 34  \\
  0.1 & 50.0 & 5.5 & 44.5 & 1.71 & 1.72 & 34 & 17  \\
  0.2 & 43.6 & 7.3 & 49.1 & 3.70 & 0.50 & 73 & 32  \\
  0.3 & 35.7 & 5.8 & 58.5 & 5.29 & 0.19 & 104 & 37  \\
  0.4 & 25.8 & 6.3 & 67.9 & 1.85 & 1.59 & 36 & 9  \\
  0.5 & 20.3 & 6.1 & 73.6 & 1.65 & 1.79 & 33 & 7  \\
\hline\hline
\end{tabular}
\end{table*}

\section{MAINTENANCE OF THE METALLIC ENHANCEMENT}

The problem of the long term preservation of the metallic excess in  the surface
stellar layers obtained by accretion is a difficult one. Let us first
consider the situation concerning the convective layers. As mentioned
in this work, the accretion of solid matter devoid of hydrogen  refers only to
the contamination of the surface convective layers. In this way, the sizes
of these layers are of a prime importance. The present majority
of SWP belongs to FGK types stars and future results on large searches
on A and M type stars will bring more light on this problem. Due to
the importance of M dwarfs as being an important part of the stellar
content of the Galaxy, planet searches are active for these stars.
Nevertheless, only few unities of M dwarfs planet hosts have been
discovered, for example, GJ876 (a multiple planet host), GJ 581 and GJ
436 (only with candidate planetary companions). even few, they are
producing interesting and instructive results. Recent detailed
spectroscopic analyses of these stars by Bean, Benedict \& Endl (2006)
and another using photometry (Bonfils et al. 2005), indicates that all
these dwarfs are metal deficient. At present it is not clear if these
results are a consequence of a special metallic distribution of the M stars
candidates for the planet searches, but independently of this, we can
consider here that due to their very large diluting convective zones,
no supersolar M dwarfs planets hosts are expected if the accretion
mechanism is working. In this way, an eventual future discovery of a
supersolar M dwarf planet host will contribute to the validity of the
primordial mechanism as mentioned in the introduction section.

Concerning the question if any metallic enhancement produced in the first 20 -
30 Myr will be maintained up to Solar ages or larger, we do not have a clear
answer yet. In fact, even if the convective zones are relatively constant from
$10^8$ up to $10^9$ yr (even diminishing somewhat near $10^{10}$ yr as can be
seen in Fig. 5 of Ford et al. 1999) other mechanisms as long term
diffusion can be acting. As suggested by Gonzales (2006), one
interesting way to investigate the
maintenance of the excess metallicity created by accretion, consists
in doing planet searches on specific open clusters. For example, in
M67 with an age and metallicity similar to that of the Sun. In these
kind of researches with open clusters, where ages and metallicities
are no more variables, any SWP
which would be enriched during the pre-main sequence would present
smaller metallicities during the sub-giant and giant phases, testing
this way the accretion mechanism.

At present, the only cluster surveyed for planet searches is the Hyades, due
to its favorable conditions as proximity and high metal abundance.
Even if no planets have been detected among their dwarf stars  
(Cochran, Hatzes \& Paulson 2002, Paulson, Cochran \& Hatzes 2004), 
the  first planet (a massive one), ever discovered in a cluster, 
has been  recently detected around a genuine
member of the Hyades (Sato et al. 2007). The host star is the
classified K0III clump giant Eps Tau (also massive) with a measured
metallicity of [Fe/H] = 0.17 $\pm$ 0.04, a value similar to that of the
Hyades cluster ([Fe/H] = 0.14 $\pm$ 0.05 Perryman et al. 1998; [fe/H] =
0.13 $\pm$ 0.01 Paulson, Sneden \& Cochran et al. 2003). Is this discovery  
a confirmation of
the primordial mechanism ? We believe that is could be a premature
conclusion because there is always the possibility that this star when
was at its pre-main sequence phase, could have been highly metallic
enriched by accretion and that this metallicity was after diluted by
convection during the successive sub-giant and giant phases.

Independent of the presence of planets, Gonzales (2006) suggests that studies
searching for trends between [Fe/H] and spectral types, as those
realized by Dotter \& Chaboyer (2003) on the Hyades stars must be
extended to other clusters. We can add that this extension can also
be applied to studies searching contaminated stars in a cluster by
their position in a H-R diagram as those realized in the Hyades by  Quillen (2002).

\section{Conclusions}

With the purpose to explore a possible mechanism of stellar self-
enrichment by an injection of rocky type planetesimals, we realized
N-body simulations in which a forced internal migration of a planet
interacts with supposed planetesimals distributed in a plane disc.
Several possibilities are considered in respect to the planets
eccentricities, migration rates and mass. Contrary to QH00 where they
considered specially the effect of 3:1 and 4:1 mean motion resonances
between the planetesimals and the planet, we found that the 2:1 mean
motion resonance is the main mechanism to drive planetesimals to the
surface of the star. This mechanism is essentially efficient for slow
migration rates of ($\tau = 10^5-10^6$ yr) and low planet
eccentricities. 

QH00 restricted themselves to study resonances distant enough from
the planet, since planetesimals with orbits that become planet-crossing 
are more likely to be ejected from the system rather than impact the star. 
However, the 2:1 resonance has a protection mechanism that avoid
close encounters between the planetesimal and the planet.
Then planetesimals locked in the 2:1 resonance are allowed to reach high 
eccentricities and impact the star. The case of $e_p=0$ was also
not considered by QH00 because for capture in the 4:1 and 3:1 resonances
it is necessary a minimum eccentricity. Nevertheless, our results (Figure 6)
show that for $e_p=0$ almost all planetesimals are captured in the 2:1 
resonance and impact the star along the planets migration  

Considering the necessary discs masses in order to provoke a
migration, we are able to set the most realistic conditions for an
enhance of the metal content of the stars. A migration of a one
Jupiter mass planet from 5 Au up to 0.01 AU requires excessively large
disc masses. The mass requirements are 10 times smaller, and became
realistic, for a migrating 0.1 Jupiter mass planet. Our simulations
with this last kind of planet and for a slow migration rate ($\tau = 10^6$
yr) stopped at pericentric distances between 0.19 and 1.79 AU when the
last planetesimals are removed from the system. Taking into account
the mass of the total impacting planetesimals we obtain maximum metal
enrichments of the order of [Fe/H]=0.18 dex. produced at ages of the
stars between 20 and 30 Myr. These results need however, primitive
disc masses three times larger than that of a primitive solar disc
nebula. These calculations open the possibility nevertheless, width
larger primitive discs (but possible to exist), to explain the metal
abundances of the observed hot Jupiters exoplanets located at ~0.05
from the stars.

\section{Acknowledgments}

 This work was partially supported by CAPES, CNPq and FAPESP.
The suggestions of an anonymous referee helped to improve this paper.

\section{References}

Adams, F.C. and Laughlin, G., 2003, Icarus, 163, 290

Armitage, P.J., 2003, ApJ, 582, L47 

Artymowicz, P., 2006
Graduate School in Astronomy: X Special Courses at the National
Observatory of Rio de Janeiro; X CCE, held 26-30 September, 2005 in
Rio de Janeiro, Brazil. AIP Conference Proceedings, Vol. 843. 
by Simone Daflon, Jailson Alcaniz, Eduardo Telles, and Ramiro de la
Reza (eds.), American Institute of Physics, Melville, 2006, p.3-34

Asphaug, E. \& Benz, W., 1996, Icarus, 121, 225

Butler, R.P. et al., 2006, ApJ 646, 505

Bazot, M., Vauclair, S., Bouchy, F., Santos, N. C., 2005, A\&A 440, 615

Bazot, M. \& Vauclair, S., 2004, A\&A 427, 695

Bean, J.L., benedict, G.F. \& Endl, M. 2006, ApJ, 653, L65

Bonfils, X. et al. 2005, A\&A, 442, 635

Cochran, W.D., Hatzes, A. P. \& Paulson, D.B. 2002, AJ, 124, 565

Davis, S.S., 2005, AJ, 627, L153

de la Reza, R. et al., 2004, 219th IAUS, 783

Dotter, A. \& Chaboyer, B. 2003, ApJ, 596, L101

Fischer, D.A. \& Valenti, J., 2005, ApJ, 622, 1102

Fogg, M.J. \& Nelson,R.P., 2005,  A\&A 441, 791

Ford, E.B., Rasio, F.A. \& Sills, A., 1999, ApJ 514, 411

Gonzalez, G., 2003, Rev. Mod. Phys, 75, 101

Gonzalez, G., 2006, PASP 118, 1494

Greaves, J.S., Fischer, D.A. \& Wyatt, M.C., 2006, MNRAS, 366, 283

Levison, H.F., Morbidelli, A., Gomes, R. \& Backman, D., 2007,
Protostars and Planets V, B. Reipurth, D. Jewitt, and K. Keil (eds.),
University of Arizona Press, Tucson, 951 pp., 2007, p.669-684

Levison, H.F. \& Duncan, M.J., 1994, Icarus, 108, 18

Lufkin, G., Richardson, D.C. \& Mundy, L.G., 2006, AJ, 653, 1464

Mandell, A.M. \& Sigurdsson, S., 2003, ApJ, 599, L111

Murray, N., et al., 1998, Science, 279, 69

Paulson, D.B., Cochran, W.D. \& Hatzes, A.P. 2004, AJ, 127, 3579

Paulson, D.B., Sneden, C. \&  Cochran, W.D. 2003, AJ, 125, 3185

Perryman, M.A.C. et al. 1998, A\&A, 331, 81

Pinz\' on, G. et al. 2007, to be submitted to A\&A

Quillen, A.C. 2002, AJ, 124, 400

Quillen, A.C. and Holman, M., 2000, AJ, 119, 397

Quillen, A. C., 2006, MNRAS, 365, 1367

Raymond, S.N., Quinn, T. \& Lunine, J.I., 2006, Icarus, 183, 265

Santos, N.C. et al., 2005, A\&A, 437, 1127

Sato, B. et al. 2007, ApJ, in press

Sicilia-Aguilar, A., et al. 2005, AJ, 129, 363

Sozzetti, A. 2004, MNRAS, 354, 1194

Sridhar, S. \& Tremaine, S. 1992, Icarus, 95, 86

Wisdom, J. \& Holman, M., 1991, AJ, 102, 1528

\end{document}